\documentclass[a4paper, 11pt]{article}
\usepackage[utf8x]{inputenc}
\usepackage[english]{babel}
\usepackage{amsfonts,amssymb, amsmath, amsthm, stmaryrd, mdframed}
\usepackage{pifont, indentfirst, dsfont, relsize}
\usepackage{fancyhdr}
\usepackage{color}
\definecolor{linkcolor}{rgb}{0,0,0.6}
\usepackage[	pdftex,colorlinks=true,	
			pdfstartview=FitV,
			linkcolor= linkcolor,
			citecolor= linkcolor,
			urlcolor= linkcolor,
			hyperindex=true,
			hyperfigures=false]
			{hyperref}

\usepackage{lastpage}
\date{}
\newcommand{\slr}{\mathfrak{sl}(2, \mathbb{R})}
\newcommand{\osp}{\mathfrak{osp}}

\newcommand{\hsl}{\mathfrak{hs}[\lambda]}
\newcommand{\shsl}{\mathfrak{shs}[\lambda]}

\newcommand{\C}{\mathcal{C}_2}
\newcommand{\W}{\mathcal{W}}
\newcommand{\Winf}{\mathcal{W}_\infty}
\newcommand{\R}{\mathbb{R}}

\newcommand{\U}{\mathcal{U}}

\newcommand{\N}{\mathbb{N}}

\newcommand{\PRS}{\text{PRS}}

\newcommand{\id}{\mathlarger{\mathds{1}}}

\let\oldsqrt\sqrt
\def\sqrt{\mathpalette\DHLhksqrt}
\def\DHLhksqrt#1#2{%
\setbox0=\hbox{$#1\oldsqrt{#2\,}$}\dimen0=\ht0
\advance\dimen0-0.2\ht0
\setbox2=\hbox{\vrule height\ht0 depth -\dimen0}%
{\box0\lower0.4pt\box2}}

\def\a{\alpha}
\def\b{\beta}

\definecolor{rougef}{rgb}{0.56,0,0}
\definecolor{vertf}{rgb}{0,0.5,0}
\definecolor{bleuf}{rgb}{0,0,0.8}

\newcommand{\email}[1]{\href{mailto:#1}{\tt{#1}}}

\newtheorem*{theorem*}{Theorem}

\newtheorem*{lemma*}{Lemma}
\theoremstyle{definition}

\newtheorem*{definition*}{Definition}

\begin{document}

\begin{center}
  \rule{.6\textwidth}{.9pt}\\
  \vspace{10pt}
  \textbf{\Large Structure constants of $\shsl\,$: \\ 
    the deformed-oscillator point of view}\\
  \rule{.6\textwidth}{.9pt}
\end{center}

\vspace*{.6cm}

\begin{center}
Thomas Basile$^{a,b,}$\footnote{E-mail address:
  \email{thomas.basile@umons.ac.be}} and {Nicolas
  Boulanger$^{a,}$\footnote{Research Associate of the Fund for
    Scientific Research$\,$-FNRS (Belgium);
    \email{nicolas.boulanger@umons.ac.be}}}
\end{center}

\vspace*{.4cm}

\begin{footnotesize} 
\begin{center}
$^{a}$
Groupe de M\'ecanique et Gravitation\\
Unit\'e de Physique Th\'eorique et Math\'ematique\\
Universit\'e de Mons -- UMONS\\
20 Place du Parc\\
7000 Mons, Belgium\\
\vspace{2mm}{\tt \footnotesize }
$^{b}$
Laboratoire de Math\'ematiques et Physique Th\'eorique\\
Unit\'e Mixte de Recherche $7350$ du CNRS\\
F\'ed\'eration de Recherche $2964$ Denis Poisson\\
Universit\'e Fran\c{c}ois Rabelais, Parc de Grandmont\\
37200 Tours, France \\
\vspace{2mm}{\tt \footnotesize }

\end{center}
\end{footnotesize}
\vspace*{.4cm}

\begin{abstract}
  \noindent
  We derive and spell out the structure constants of the
  $\mathbb{Z}_2$-graded algebra $\shsl\,$ by using
  deformed-oscillators techniques in $Aq(2;\nu)\,$, the universal
  enveloping algebra of the Wigner-deformed Heisenberg algebra in 2
  dimensions. The use of Weyl ordering of the deformed oscillators is
  made throughout the paper, via the symbols of the operators and the
  corresponding associative, non-commutative star product. The
  deformed oscillator construction was used by Vasiliev in order to
  construct the higher spin algebras in three spacetime dimensions. We
  derive an expression for the structure constants of $\shsl\,$ and
  show that they must obey a recurrence relation as a consequence of
  the associativity of the star product.  We solve this condition and
  show that the $\hsl\,$ structure constants are given by those
  postulated by Pope, Romans and Shen for the Lone Star product.
\end{abstract} 

\thispagestyle{empty}

\newpage
\tableofcontents

\setcounter{page}{1}

\section{Introduction} \label{sec:intro}
Three dimensional spacetime constitutes a particularly interesting
testing ground for the study of Higher Spin (HS) theories, as some of
the technical difficulties appearing in dimensions 4 and higher are
absent due to the topological nature of gauge fields of spin 2 and
higher. \\

In the presence of a negative cosmological constant in 3D and without
any coupling between HS fields and matter, a standard action principle
for an infinite tower of gauge fields with integer spin $s\geqslant 2$
is given \cite{Blencowe:1988gj} by (the difference of) two
Chern--Simons actions for the infinite dimensional superalgebras
studied by Fradkin and Vasiliev, see
\cite{Vasiliev:1986qx,Fradkin:1986ka}.  A doubling of the algebra,
necessary in order to produce the direct sum in the three-dimensional
anti-de Sitter (AdS$_3$) isometry algebra $\mathfrak{so}(2,2) \cong
\mathfrak{so}(2,1) \oplus \mathfrak{so}(2,1)\,$, can be achieved by
introducing an outer element $\psi\,$ satisfying $\psi^2=1\,$.  As is
customary in this context, one does not always mention this doubling
of algebras that is implicitly understood.  As shown in
\cite{Vasiliev:1989re}, these (super)algebras are characterized by a
real parameter $\nu$ and possess a unique nondegenerate
supertrace\,\footnote{The existence of a one parameter family of
  higher spin algebras is usually considered to be a special feature
  of dimension 3, but it is not the case, see
  \cite{Fradkin:1990ir,Fernando:2009fq} on AdS$_5$ and
  \cite{Boulanger:2011se,Joung:2014qya} in general dimension.}.  For
critical values $\nu = -(2\ell + 1)\,$, $\ell \in \mathbb{N}\,$, an
ideal appears and can be quotiented out, leaving a finite-dimensional
bosonic subalgebra $\mathfrak{gl}(N)\,$ where $N = \ell+1\,$, see
e.g. \cite{Boulanger:2013naa} for a recent review and extensions,
together with the notations that we adopt here.\footnote{The
  Chern--Simons construction in \cite{Boulanger:2013naa} unifies HS
  fields with fractional spin-fields and an internal nonabelian
  sector.}  In the critical case, the $\mathfrak{sl}(N)\oplus
\mathfrak{sl}(N)$-valued connection describes gauge fields with spin
$s=2, 3, \ldots, N\,$, thereby naturally extending the classical
reformulation of three-dimensional gravity \cite{Achucarro:1987vz,
  Witten:1988hc}, see \cite{Campoleoni:2010zq} for a pedagogical
review.  The Blencowe construction \cite{Blencowe:1988gj} corresponds
to taking $\nu = 0\,$.\\

Focusing on the purely bosonic case, the HS gauge algebras that
generalise the one used by Blencowe are denoted $\hsl\,$ and defined
as follows:
\begin{equation}
  \begin{aligned}
    \mathbb{C} \oplus \hsl   = \frac{\U \left( \slr \right)}{\langle \C
      - \mu \, \id \rangle}, & \ & \quad \mu = \frac{\lambda^2 - 1}4\;,
  \end{aligned}
  \label{hsl_def}
\end{equation}
where $\U \left( \slr \right)$ is the universal enveloping algebra
(UEA) of $\slr$, $\C$ its quadratic Casimir and $\langle \C - \mu \id
\rangle$ the ideal generated by the relation in brackets, i.e. the
value of $\C$ is fixed. The relation between the parameters $\nu$ and
$\lambda$ is given by $\lambda = \frac{1 - \nu}{2}\,$.  When the value
of $\nu$ is non critical, the Chern--Simons model describes an
infinite tower of non-propagating but interacting HS fields of spin
$s=2,3,\ldots\,$. \\

The Prokushkin--Vasiliev (PV) equations \cite{Prokushkin:1998bq}
precisely describe an infinite tower of higher spin gauge fields,
coupled with two (or one) complex scalar field.  As in the
Chern--Simons formulation, the PV equations are based on the gauging
of $\hsl\,$.  The analysis of the spectrum of the PV equations is not
straightforward however, due to the presence of extra kleinians and
twisted sectors; see e.g. \cite{Kessel:2015kna, Arias:2016ajh} for
recent studies and in particular \cite{Kessel:2015kna} for a thorough
treatment of the twisted sector.  Even though in the Blencowe
formulation one benefits from the standard technology from
Chern--Simons actions, matter coupling is not possible, or at least it
is not yet known how to implement it. For some works in that
direction, see \cite{Fujisawa:2013ima}.  Note, however, that it is
possible to give an action for matter-coupled 3D HS fields, see
\cite{Bonezzi:2015igv}. There, it was shown how the Chern--Simons
description emerges upon consistent truncation and at the expense of
losing the matter sector.  In \cite{Bonezzi:2015igv,Arias:2016ajh} it was
also shown how to further truncate the spectrum to only one \it real
\rm scalar field, of mass $m^2 = -1 + \lambda^2\,$.
This last feature may be interesting in the context of the work 
\cite{Gaberdiel:2011nt} and after the more recent developments where
it was understood that one of the two scalar fields should correspond
to non-perturbative degrees of freedom, see the review
\cite{Gaberdiel:2012uj}. \\

The PV model has witnessed a surge of interest from the fact that the
bulk theory governed by the PV equations has been conjectured by
Gaberdiel and Gopakumar \cite{Gaberdiel:2010pz, Gaberdiel:2012uj} to
be dual to $\Winf[\lambda]$ minimal model CFTs.  More precisely, the
CFT considered is a Wess--Zumino--Witten coset model:
\begin{equation}
  \frac{\text{SU}(N)_k \otimes \text{SU}(N)_1}{\text{SU}(N)_{k+1}}
\end{equation}
in the t'Hooft limit 
\begin{equation}
  N, k \rightarrow \infty, \ \text{ with } \ \lambda = \frac{N}{N+k}
  \ \text{fixed.}
\end{equation}
The t'Hooft parameter $\lambda$ is to be identified with the parameter
fixing the $\slr$ quadratic Casimir value in the definition of
$\hsl\,$.  Promising results in this holographic context have been
obtained \cite{Gaberdiel:2014cha, Gaberdiel:2015mra,
  Gaberdiel:2015wpo}, in favour of HS theories seen as a tensionless
limit of String Theory, see refs. therein. \\

A realisation of $\hsl\,$ was given in PV's original paper
\cite{Prokushkin:1998bq}, elaborating on the previous work of Vasiliev
\cite{Vasiliev:1989re} using the so-called \emph{Wigner-deformed
  oscillators} \cite{Wigner:1950,Yang:51}, see also
\cite{Plyushchay:1994re,Plyushchay:1997ty}, while \cite{Joung:2014qya}
gives a matrix-valued realisation of the deformed oscillators.
They constitute a deformation of the usual oscillators as they verify
the following commutation relation:
\begin{equation}
  \begin{aligned}
    \left[ \hat q_\a, \hat q_\b \right] = 2i \, \epsilon_{\a \b} \, (
    1 + \hat k\nu ), & \ & \{ \hat q_\a, \hat k \} = 0, & \ & \nu \in
    \R
  \end{aligned}
  \label{def_osc_alg}
\end{equation}
in terms of a real spinor $\hat q_\a$ ($\alpha = 1,2$) of $\slr\,$ and
the Klein operator (or kleinian) $\hat{k}\,$ obeying
$\hat{k}^2=\id\,$.
The matrix of components $\epsilon_{\alpha\beta} = -
\epsilon_{\beta\alpha}$ is given by $\epsilon_{12}=1\,$.
The above deformed oscillators can indeed be used to realise $\slr
\cong \mathfrak{so}(1,2) \cong \mathfrak{sp}(2, \R)\,$ by defining the
generators
\begin{equation}
  T_{\a \b} = \tfrac{1}{4i} \{ \hat q_\a, \hat q_\b \}
  \label{def_gen_slr}
\end{equation}
that obey  
\begin{equation}
  \begin{aligned}[]
    [ T_{\a \b}, T_{\gamma \delta} ] & = 2 \left( \epsilon_{\a
      (\gamma}T_{\delta) \b} + \epsilon_{\b (\gamma}T_{\delta) \a}
    \right) \ ,
  \end{aligned}
\end{equation}
where indices inside brackets are symmetrised with strength one.  The
deformed oscillators can also be used to present $\osp(2 \rvert 2)$
upon defining
\begin{equation}
  \begin{aligned}
    Q_\a^{(1)} := \hat q_\a, & \ & Q_\a^{(2)} := \hat k \,\hat q_\a, &
    \ & J := \tfrac{1}2 (\hat k+\nu) \, ,
  \end{aligned}
\end{equation}
yielding \cite{Bergshoeff:1991dz}
\begin{equation}
  \begin{array}{cclcccl}
    [T_{\a \b}, Q_\gamma^{(i)}] & = & - 2\epsilon_{\gamma (\alpha}
    Q_{\b)}^{(i)} \, , & \ & \{ Q_\a^{(i)}, Q_\b^{(j)} \} & = & 4i
    \left( \sigma_3^{ij}\, T_{\a \b} - \tau^{ij}\, \epsilon_{\a \b} J
    \right)\, , \\ \left[T_{\a \b}, J \right] & = & 0 \, , & \ & [J,
      Q^{(i)}_\a] & = & \tau^{ij} Q^{(j)}_\a \, ,
  \end{array}
  \label{osp22}
\end{equation}
with $\tau^{ij} = -\tau^{ji}$, $\tau^{12} =1\,$; and $\sigma_3$ is the
third Pauli matrix. Both restricted set of generators 
$\left\{T_{\a \b}, Q_\a^{(i)} \right\}$, with $i$ equals to 1 or 2, 
span an $\osp(1 \vert 2)$ subalgebra in $\osp(2 \rvert 2)\,$.
An advantage using deformed oscillators is that they automatically enforce 
the quotient in \eqref{hsl_def}, as well as:
\begin{equation}
  \mathbb C \oplus \shsl = \frac{\U(\osp(1 \rvert 2))}{\langle \C -
    \tfrac{1}{4}\lambda(\lambda-1)\id \rangle}\,,
  \label{shsl_def}
\end{equation}
defining the fermionic extension of $\hsl\,$, containing generators of 
all (half-)integer spins $s \geqslant 3/2$ and where $\C$ here denotes 
the $\osp(1 \rvert 2)$ quadratic Casimir. \\

The higher spin algebra is $\hsl$ is obtained by considering the
commutators in the associative algebra made out of all possible \it
even \rm powers of the deformed oscillators, endowed with the
associative \it deformed\,\footnote{The terminology ``deformed star
  product'' refers to both the fact that this product is defined on
  the deformed oscillators, but also to the fact that, upon setting
  the parameter $\nu$ to 0, one recovers the usual Moyal star
  product.} \rm star product:
\begin{equation}
  q_\a \star q_\b = q_\a q_\b + i\epsilon_{\a \b} (1 + k\nu)\ ,
  \label{WeylOrder}
\end{equation}
where now $q_\a$ and $k$ (without hats) are the so-called {\it
  symbols} associated with the operators $\hat q_\a$ and $\hat k\,$.
Very schematically, the main idea of the symbol calculus goes as
follows.  To any operator $\hat{A}(\hat{q},\hat{k})\,$ one associates
a classical function $A_{\cal O}({q}_\alpha,{k}) \equiv
[\hat{A}(\hat{q},\hat{k})]_{\cal O}\,$ obtained by (i) first ordering
the operators $\hat{q}_\alpha$ and $\hat{k}$ entering the expression
of $\hat{A}(\hat{q},\hat{k})\,$ by following a given ordering
prescription ${\cal O}$ and making use of the relations
\eqref{def_osc_alg}, and (ii) by then replacing the operators
$\hat{q}_\alpha$ and $\hat{k}$ by the classical, i.e. commuting
\emph{symbols} $q_\alpha$ and $k\,$.  The composition of operators
ordered according to the prescription ${\cal O}$ is represented, in
the (appropriately defined) space of classical symbols, by an
associative but noncommutative star-product $\star_{\cal O}$ such that
\begin{equation}
  [ \hat{A}\, \hat{B} ]_{\cal O} \,=\, A_{\cal O} \;\star_{\cal O}
  \;B_{\cal O} \;.
\end{equation}
For more precise statements and references, see
e.g. \cite{zachos2005quantum} and \cite{Iazeolla:2008ix} in the
context of higher-spin theories.  The ordering prescription we will be
considering here is the Weyl ordering, whereby all the operators
$\hat{q}_\alpha$ are symmetrised before being replaced by their
classical, commuting symbols $q_\alpha\,$. The above formula
\eqref{WeylOrder} therefore defines the star product in the Weyl
ordering prescription, where the right-hand side corresponds to the
sum of the anticommutator with the commutator of the operators
$\hat{q}_\alpha$ and $\hat{q}_\beta\,$.  As it is customary in the
present context, before replacing the operators by their symbols, one
chooses to place the kleinian $\hat{k}$ either to the left or to the
right of any monomial in the $\hat{q}_\alpha$'s.  In this paper, we
will always place $\hat k$ to the left.\\

The set of arbitrary Weyl-ordered monomials in the deformed
oscillators $q_\alpha$ together with the identity (the monomial of
degree zero) and $k$ gives a basis of the universal enveloping algebra
of $\mathfrak{osp}(1|2)\,$~\cite{Vasiliev:1989re}.  This is an
associative algebra, by construction. If one endows it with the
star-commutator, one gets a Lie (super)algebra denoted
$\mathfrak{shs}[\lambda]\,$ \cite{Fradkin:1986ka} where the
$\mathbb{Z}_2$ grading is given by the order of a monomial in the
$q_\alpha$, modulo 2 \cite{Vasiliev:1989re}.
The structure constants of the algebra $\hsl$ were conjectured to be
given by the commutator lone-star product of
\cite{Pope:1989sr,Pope:1990kc,Pope:1991ig,Fradkin:1990qk}, see more
recently the appendix B of \cite{Ammon:2011ua} where some evidences
for this conjecture were given up to spin 4.\\

The present paper aims at explicitly computing the structure constants
appearing in the star product of two arbitrary (even or odd)
Weyl-ordered monomials in the deformed oscillators.  In particular, we
will \emph{prove} that the lone-star product produces the correct
structure constants for the bosonic sector of the associative algebra
$Aq(2;\nu)$ underlying $\hsl\,$, and this without making any
restriction on the value of the spin of the generators in involved.
What we understood of the interesting recent work
\cite{Korybut:2014jza} is that the lone-star product was assumed to be
the underlying product for $\hsl\,$ and was extended to $Aq(2;\nu)$ by
using associativity, which the author then proved in the appendix in
\cite{Korybut:2014jza}.  In other words, by conjecturing the lone-star
product for the algebra $Aq(2;\nu)$, the author proved the
associativity property.  In the present paper, we proceed differently
by starting from the associative algebra $Aq(2;\nu)\,$ and showing
that the lone-star product formula is the \emph{unique} solution for
the structure constants.  \\

Our paper is organised as follows:
In \hyperref[sec:2]{Section 2}, we derive a few lemmas on the star
product of deformed oscillators and use them to obtain a closed
formula for the searched-for structure constants.  The
\hyperref[sec:3]{Section 3} consists of a brief review of $\W$
algebras and their connections to (three dimensional) higher spin
theories.  This connection is exploited in \hyperref[sec:4]{Section 4}
to derive the structure constants of $\shsl\,$. In the bosonic
restriction, the structure constants can be rewritten in terms of
(generalised) hypergeometric functions, as postulated by Pope, Romans
and Shen \cite{Pope:1989sr} in a different context.
%

\section{Some deformed oscillator algebra} \label{sec:2}
%
After recalling some basic facts about the deformed oscillators
algebra, we proceed in the following section to prove some lemmas that
will be needed in order to derive the structure constant of $\hsl$.

\subsection{Simple (anti)commutators.}
\noindent Taking as a starting point the deformed oscillator
star-commutation relation:
\begin{equation}
  \left[ q_{\a}, q_{\b} \right]_\star \;=\; 2{\rm{i}}\,\epsilon_{\a\b}
  \;(1+\nu\,k)\quad, \qquad \left\{ k, q_{\a} \right\}_\star=0 \ ,
\end{equation}
we would like to compute the star product of $q_{\a}$ with the
completely symmetrised (\emph{i.e.} Weyl-ordered) product
\begin{equation}
  q_{\b_1}\ldots q_{\b_n} \; :=\;
  \frac{1}{n!}\;\sum_{\sigma\in\mathfrak{S}_n}
  q_{\b_{\sigma(1)}}\star\, q_{\b_{\sigma(2)}}\star\, \ldots\star \,
  q_{\b_{\sigma(n)}}\quad,
\end{equation}
where $\mathfrak{S}_n$ denotes the group of permutations of $n$
elements. The usual decomposition rule (Pieri's rule) for the tensor
product of irreducibles of $\mathfrak{S}_n$ into irreducibles of
$\mathfrak{S}_n$ gives
\begin{eqnarray}
  q_{\a}\star(q_{\b_1}\ldots q_{\b_n}) &=& q_{\a} q_{\b_1}\ldots
  q_{\b_n} +
  {\frac{2}{(n+1)}\;\frac{1}{n!}}\;\sum_{\sigma\in\mathfrak{S}_n}
  \left[ \stackrel{\ulcorner}{ q_{\a}}\star\,
    \stackrel{\urcorner}{q_{\b_{\sigma(1)}}}\star\,
    q_{\b_{\sigma(2)}}\ldots\star\, q_{\b_{\sigma(n)}} + \right.
    \label{3}\\ && + \left.
    \stackrel{\ulcorner}{q_{\a}}\star\,q_{\b_{\sigma(2)}}\star\,
    \stackrel{\urcorner}{q_{\b_{\sigma(1)}}}\star\,\ldots\star\,
    q_{\b_{\sigma(n)}}\;+\;\ldots\; +\;
    \stackrel{\ulcorner}{q_{\a}}\star\,q_{\b_{\sigma(2)}}\star\,
    \ldots\star q_{\b_{\sigma(n)}}
    \star\,\stackrel{\urcorner}{q_{\b_{\sigma(1)}}}\right] \nonumber
\end{eqnarray}
where we introduced the following notation for antisymmetrization:
\begin{equation}
  \stackrel{\ulcorner}{a_1} a_2 \ldots \stackrel{\urcorner}{a_i}
  a_{i+1}\ldots a_n = \frac{1}{2}\;\Big( a_1 a_2\ldots a_{i-1}\,a_i
  \,a_{i+1}\ldots a_n \;-\; a_i \,a_2\ldots a_{i-1}a_1 \,a_{i+1}\ldots
  a_n\Big)\quad.
\end{equation}
In Eq. (\ref{3}), it is possible to bring next to each other every two
elements $q_{\a}$ and $q_{\b_{\sigma(1)}}$ that are antisymmetrized in
the sum, so as to produce a commutator
${[}q_{\a}\,,q_{\b_{\sigma(1)}}{]}_\star\,$.  This can be done by
dragging a term sitting at the place $i$ in the chain of star product
to the place $2$, thereby producing extra terms with commutators.
Summing up everything together, one finds
\begin{eqnarray}
  q_{\a}\star(q_{\b_1}\ldots q_{\b_n}) &=& q_{\a}q_{\b_1}\ldots
  q_{\b_n} + {\frac{1}{n!}}\; \sum_{\sigma\in\mathfrak{S}_n}
  \frac{2{\rm{i}}}{(n+1)} \times \;
  \label{4}\\
&& \times\left[ \sum_{i=0}^{n-1}
    (n-i)\,q_{\b_{\sigma(1)}}\star\,\ldots\star
    q_{\b_{\sigma(i)}}\star(1+\nu\,k)\epsilon_{\a\b_{\sigma(i+1)}}
    \star\,q_{\b_{\sigma(i+2)}}\star\ldots\star q_{\b_{\sigma(n)}}
    \right]\;.\quad\, \nonumber
\end{eqnarray}
At this stage, it is a matter of performing the sums corresponding to
the ``1'' and to the ``$\nu k$'' in the $(1+\nu\,k)$'s appearing at
the different places in the chain of star product.  The first sum is
easy to do: $\sum_{i=0}^{n-1}(n-i)=\frac{n(n+1)}{2}\;$.  One has to be
more careful with the second one (involving the kleinian $k$) since it
produces an alternating sum and a distinction must be done between the
cases where $n$ is even and odd. It is nevertheless straightforward
and gives the final answer
\begin{equation}
  q_{\a}\star(q_{\b_1}\ldots q_{\b_n}) \;=\; q_{\a}q_{\b_1}\ldots
  q_{\b_n} \;+\; {\rm{i}}\,n\,\left(
  1+\frac{2n+1-(-1)^n}{2n(n+1)}\;\nu\,k \right)
  \epsilon_{\a(\beta_1}q_{\b_2}\ldots q_{\b_n)}
\end{equation}
that correctly reproduces the case $n=1\,$. Similarly one gets
\begin{equation}
  (q_{\b_1}\ldots q_{\b_n})\star q_{\a} \;=\; q_{\a} q_{\b_1}\ldots
  q_{\b_n} \;-\; {\rm{i}}\,n\,\left( 1 +
  (-1)^{n+1}\frac{2n+1-(-1)^n}{2n(n+1)}\;\nu\,k \right)
  \epsilon_{\a(\beta_1} q_{\b_2}\ldots q_{\b_n)}\quad.
\end{equation}
Using these two results, one finds the following commutators and
anticommutators:
%
\begin{eqnarray}
    \left[ q_\alpha \, ,\, (q_\beta)^n \right]_\star &=& 2
    {\rm{i}}\epsilon_{\alpha\beta} \, (n + k\nu \,P_{n})\;
    (q_\beta)^{n-1}\,\quad, \qquad \quad P_n :=
    \tfrac{1-(-1)^n}{2}\;
    \label{comm1n}
    \\ \left\{ q_\alpha \, ,\, (q_\beta)^n \right\}_\star &=& 2\,
    q_\alpha (q_\beta)^n + 2 {\rm{i}}\epsilon_{\alpha\beta}\,
    \tfrac{n}{(n+1)}\; k\nu \,P_{n+1}\; (q_\beta)^{n-1} \quad .
    \label{anticomm1n}
\end{eqnarray}
\noindent As a corollary, we find
\begin{eqnarray}
  \left[q_\a , (q_\a)^r (q_\b)^s \right]_\star &=&
  2{\rm{i}}\epsilon_{\a\b}\,{\tfrac{s}{(r+s)}}\;\left[(r+s) + k\nu \,
    P_{r+s}\right] \,(q_\a)^r(q_\b)^{s-1}\quad , \label{mixed_com}
  \\ \left\{ q_\a , (q_\a)^r (q_\b)^s \right\}_\star &=& 2\,
  (q_\a)^{r+1}(q_\b)^{s} + (2{\rm{i}}\epsilon_{\a\b})\,k\nu\,
  \,{\tfrac{s}{(r+s)}}\;\tfrac{(r+s)}{(r+s+1)}\; P_{r+s+1} \,(q_\a)^r
  (q_\b)^{s-1}\;.
  \label{mixed_anticom}
\end{eqnarray}
We use notation whereby repeated indices are completely
symmetrized with strength one, and $(q_\beta)^n$ stands for
$q_{\beta_1}\ldots q_{\beta_n}\,$.
Grouping together the commutator \eqref{comm1n} and anticommutator
\eqref{anticomm1n} , one derives the following formula which is
central in the forthcoming computations involving deformed
oscillators:
\begin{eqnarray}
    q_\alpha \, \star\, (q_\beta)^n & = & q_\alpha (q_\beta)^n +
    {\rm{i}}\epsilon_{\alpha\beta} \, n \,Y^+_n \, (q_{\beta})^{n-1}\ ,
    \quad \qquad Y^{\pm}_n \; := \; 1 \pm k\nu \,\left[ \tfrac{P_n}n +
      \tfrac{P_{n+1}}{n+1} \right] \ , \qquad \label{nest1}
    \\ q_\alpha \, \star\, [ (q_{\a})^r (q_{\beta})^s ] & = &
    (q_\alpha)^{r+1} (q_\beta)^s + {\rm{i}}\epsilon_{\alpha\beta} \,
    s\; Y^+_{r+s}\, (q_{\alpha})^r
    (q_{\beta})^{s-1}\ \quad. \label{nest2}
\end{eqnarray}
A similar relation can be derived for the star product of one oscillator
with a monomial from the right:
\begin{equation}
    \begin{aligned}
      (q_\a)^n \, (q_\b)^m \star q_\b = (q_\a)^n \, (q_\b)^{m+1} +
      i\epsilon_{\a \b}\, n\, \bar Y_{n+m} (q_\a)^{n-1} \, (q_\b)^m, &
      \, & \bar Y_n = 1 + k\nu \left[ \frac{P_n}{n}
        -\frac{P_{n+1}}{n+1} \right]\ .
    \end{aligned}
    \label{right}
\end{equation}

\subsection{Recurrence relation}
\noindent By nesting formulas \eqref{nest1} and \eqref{nest2}, it is
now straightforward to compute the star-product of two monomials of
arbitrary degrees $m$ and $n$, where without loss of generality one
chooses $m\leqslant n\,$:
\begin{eqnarray}
  (q_\alpha)^m \star (q_\beta)^n &=& (q_\alpha)^{m-1} \star \left[
    q_{\alpha} (q_{\beta})^n + ({\rm{i}}\,\epsilon_{\a\b}) \, n \,
    Y_n^+ (q_{\beta})^{n-1} \right] \\ &=& (q_\alpha)^{m-2}
  \star \Big( (q_{\alpha})^2 (q_{\beta})^n +
  ({\rm{i}}\,\epsilon_{\a\b}) \tfrac{n}{(n+1)}\, (n+1) \, Y_{n+1}^+\,
  q_{\alpha}(q_{\beta})^{n-1} + \nonumber \\ && \qquad\qquad
  \qquad ({\rm{i}}\,\epsilon_{\a\b}) \, n \, Y_{n}^- \left[
    q_{\alpha}(q_{\beta})^{n-1} + ({\rm{i}}\,\epsilon_{\a\b}) \, (n-1)
    \, Y_{n-1}^+ (q_{\beta})^{n-2} \right] \Big) \nonumber = \ldots
  \label{arb_prod}
\end{eqnarray}
where one continues $(m-2)$ times until one has exhausted the degree
of the first monomial. Writing the above star product
as\,\footnote{Notice that in the general case, the upper bound of the
sum in \eqref{decomposition} is $\text{min}(m,n)$.}:
\begin{equation}
  (q_\alpha)^m \star (q_\beta)^n = \sum_{p=0}^m
  \frac{(i\epsilon_{\alpha \beta})^p \, n!}{(n-p)!} \, b_p^{(m,n)}
  (q_\a)^{m-p} (q_\b)^{n-p} \ ,
  \label{decomposition}
\end{equation}
our problem then boils down to computing the structure constants
$b_p^{(m,n)}$.
\paragraph{From the left.}
To do so, we will start by using the associativity of the star product
to extract a recurrence relation on the coefficients
$b_p^{(m,n)}$. Defining $\pi(\sum_k\prod_i
Y_{n_{k,i}}^{\epsilon_{k,i}}) = \sum_k \prod_i
Y_{n_{k,i}}^{-\epsilon_{k,i}}$ (all signs of the $Y^\pm_n$ symbols are
flipped, i.e. $k\nu \rightarrow -k\nu$), we have:
\begin{eqnarray}
  (q_\a)^{m+1} \star (q_\b)^n & = & q_\alpha \star (q_\a)^m \star
  (q_\b)^n = q_\alpha \star \sum_{p=0}^m \frac{(i\epsilon_{\alpha
      \beta})^p \, n!}{(n-p)!} \, b_p^{(m,n)} (q_\a)^{m-p}
  (q_\b)^{n-p} \nonumber \\ 
  \;=\; \sum_{p=0}^m & &\!\!\!\!\!\!\!\!\!\!\!\!\!
  \frac{(i\epsilon_{\alpha \beta})^p \, n!}{(n-p)!} \,
  \pi(b_p^{(m,n)}) \Big((q_\a)^{m-p+1} (q_\b)^{n-p} +
  (i\epsilon_{\alpha \beta})(n-p)Y_{n+m-2p}^+ (q_\a)^{m-p}
  (q_\b)^{n-p-1} \Big) \nonumber \\ & = & \sum_{p=0}^{m}
  \frac{(i\epsilon_{\alpha \beta})^p \, n!}{(n-p)!} \,
  \pi(b_p^{(m,n)}) (q_\a)^{m-p+1} (q_\b)^{n-p} \nonumber \\ &
  & \hspace{.2\textwidth} + \sum_{p=1}^{m+1} \frac{(i\epsilon_{\alpha
      \beta})^p \, n!}{(n-p)!} \, \pi(b_{p-1}^{(m,n)})
  Y_{n+m-2(p-1)}^+ (q_\a)^{m+1-p} (q_\b)^{n-p} \nonumber \\ & = &
  \sum_{p=0}^{m+1} \frac{(i\epsilon_{\alpha \beta})^p \, n!}{(n-p)!}
  \, b_p^{(m+1,n)} (q_\alpha)^{m+1-p} (q_\beta)^{n-p}\ ,
\end{eqnarray}
  \begin{equation}
    \Rightarrow 
    \left\lbrace
    \begin{aligned}
      b_0^{(m,n)} & = 1, \ \forall \, m,n \in \N\, , \qquad
      b_m^{(m,n)} = \prod_{i=0}^{m-1} Y_{n-i}^{(-1)^{m+1+i}},
      \ \forall m,n \in \N \\ b_p^{(m+1,n)} & = \pi(b_p^{(m,n)}) +
      Y_{n+m-2(p-1)}^+ \pi(b_{p-1}^{(m,n)}), \ \forall \, m,n \in \N,
      \ \forall \, p \in \llbracket 1, m \rrbracket
    \end{aligned}
    \right.
    \label{recurrence} \ .
  \end{equation}
%
\vspace*{.5cm}

Notice that, in the special case $\nu = 0\,$, this recurrence relation
reduces to the one obeyed by the binomial coefficients:
\begin{equation}
  \binom{m+1}{p} = \binom{m}{p} + \binom{m}{p-1} \, .
\end{equation}
Indeed, for $\nu = 0\,$, $\pi(b_p^{(m,n)}) = b_p^{(m,n)}$ and $Y_n^\pm
= 1\,$.  This was to be expected, as one should recover the Moyal star
product, for which the coefficients of $(i \epsilon_{\alpha\beta})^p$
in \eqref{decomposition} is
$\frac{1}{p!}\,\frac{n!}{(n-p)!}\,\frac{m!}{(m-p)!}\,$.  The factor
$\frac{1}{p!}\,$ comes from the Taylor coefficient of the exponential
in the Moyal star-product formula, while the other two contributions
$\frac{m!}{(m-p)!}\,$ and $\frac{n!}{(n-p)!}\,$ come from taking $p$
derivatives of the monomials of order $m$ and $n\,$, respectively.  \\

After some brute force computation, one can guess from several
examples the following general formula:
%
  \begin{equation}
    b_p^{(m,n)} \;=\; \prod_{\ell = 0}^{p-1} \;\underset{i_0 =
      0}{\sum_{i_\ell = i_{\ell-1}}^{m-p}} Y_{n+i_\ell -
      \ell}^{(-1)^{m+i_\ell+\ell+1}}\ ,
    \label{formula}
  \end{equation}
%
or in a less compact way:
\begin{equation}
  b_p^{(m,n)} = \sum_{i_0 = 0}^{m-p} \sum_{i_1 = i_0}^{m-p} \dots
  \sum_{i_{p-1} = i_{p-2}}^{m-p} Y_{n+i_0}^{(-1)^{m+i_0+1}}
  Y_{n+i_1-1}^{(-1)^{m+i_1+2}} \dots Y_{n+i_{p-1}
    -(p-1)}^{(-1)^{m+i_{p-1}+p}}\;.
\end{equation}
This formula reproduces the coefficients calculated above, and more
importantly verifies the recurrence relation \eqref{recurrence}.
\begin{proof}
  Let us first write:
  \begin{equation}
    \bullet \hspace{10pt} \pi(b_p^{(m,n)}) = \sum_{i_0 = 0}^{m-p}
    \sum_{i_1 = i_0}^{m-p} \dots \sum_{i_{p-1} = i_{p-2}}^{m-p}
    Y_{n+i_0}^{(-1)^{m+i_0+2}} Y_{n+i_1-1}^{(-1)^{m+i_1+3}} \dots
    Y_{n+i_{p-1} -(p-1)}^{(-1)^{m+i_{p-1}+p+1}}\; ;
  \end{equation}
  \begin{equation}
    \bullet \hspace{10pt} Y_{n+m-2(p-1)}^+ \pi(b_{p-1}^{(m,n)}) =
    Y_{n+m-2(p-1)}^+ \sum_{i_0=0}^{m-p+1} \sum_{i_1 = i_0}^{m-p+1}
    \dots \sum_{i_{p-2} = i_{p-3}}^{m-p+1} Y_{n+i_0}^{(-1)^{m+i_0+2}}
    \dots Y_{n+i_{p-2} -(p-2)}^{(-1)^{m+i_{p-2}+p}}\;.
  \end{equation}
  Noticing that $Y_{n+m-2(p-1)}^+ = Y_{n+i_{p-1}
    -(p-1)}^{(-1)^{m+i_{p-1}+p+1}} \delta_{i_{p-1}, m-p+1}$, and
  denoting by the symbol $\Pi_{p-1}$ the product
  $Y_{n+i_0}^{(-1)^{m+i_0+2}} Y_{n+i_1-1}^{(-1)^{m+i_1+3}} \dots
  Y_{n+i_{p-1} -(p-1)}^{(-1)^{m+i_{p-1}+p+1}}$, the sum of the two
  terms above reads:
  \begin{equation}
    \pi(b_p^{(m,n)}) + Y_{n+m-2(p-1)}^+ \pi(b_{p-1}^{(m,n)}) =
    \Bigg(\sum_{i_0=0}^{m-p} \dots \sum_{i_{p-1} = i_{p-2}}^{m-p} +
    \sum_{i_0=0}^{m-p+1} \dots \sum_{i_{p-2} = i_{p-3}}^{m-p+1}
    \delta_{i_{p-1}, m-p+1} \Bigg) \Pi_{p-1}\;.
  \end{equation}
  On the other hand, we have:
  \begin{eqnarray}
    b_p^{(m+1,n)} & = & \sum_{i_0 = 0}^{m-p+1} \sum_{i_1 =
      i_0}^{m-p+1} \dots \sum_{i_{p-1} = i_{p-2}}^{m-p+1}
    Y_{n+i_0}^{(-1)^{m+i_0+2}} Y_{n+i_1-1}^{(-1)^{m+i_1+3}} \dots
    Y_{n+i_{p-1} -(p-1)}^{(-1)^{m+i_{p-1}+p+1}} \nonumber \\ & =
    &\Bigg( \sum_{i_0=0}^{m-p+1} \sum_{i_1 = i_0}^{m-p+1} \dots
    \sum_{i_{p-1} = i_{p-2}}^{m-p+1} \Bigg) \Pi_{p-1} \nonumber\\ & = &
    \Bigg(\sum_{i_0 = 0}^{m-p} \sum_{i_1 = i_0}^{m-p+1} \dots
    \sum_{i_{p-1} = i_{p-2}}^{m-p+1} + \delta_{i_0,m-p+1} \dots
    \delta_{i_{p-1}, m-p+1} \Bigg) \Pi_{p-1} \nonumber \\ & = & 
    \Bigg(
    \sum_{i_0 = 0}^{m-p} \sum_{i_1 = i_0}^{m-p} \sum_{i_2 =
      i_1}^{m-p+1} \dots \sum_{i_{p-1} = i_{p-2}}^{m-p+1} + \nonumber \\ & \qquad + & 
    \underbrace{\sum_{i_0=0}^{m-p} \delta_{i_1, m-p+1} \dots
      \delta_{i_{p-1}, m-p+1} + \delta_{i_0,m-p+1} \dots
      \delta_{i_{p-1}, m-p+1}}_{= \sum_{i_0=0}^{m-p+1}
      \delta_{i_1,m-p+1} \dots \delta_{i_{p-1}, m-p+1}} \Bigg)
    \Pi_{p-1} \nonumber \\ & = & \dots = \Bigg(\sum_{i_0=0}^{m-p}
    \sum_{i_1=i_0}^{m-p} \dots \sum_{i_{p-1}=i_{p-2}}^{m-p} +
    \sum_{i_0=0}^{m-p+1} \dots \sum_{i_{p-2} = i_{p-3}}^{m-p+1}
    \delta_{i_{p-1}, m-p+1} \Bigg) \Pi_{p-1} \nonumber \\ & = &
    \pi(b_p^{(m,n)}) + Y_{n+m-2(p-1)}^+ \pi(b_{p-1}^{(m,n)})\;,
  \end{eqnarray}
  i.e. the coefficients given by \eqref{formula} verify the recurrence
  relation \eqref{recurrence}.
\end{proof}

\paragraph{Reduced formula.}
Although formula \eqref{formula} is exact, it is almost impossible to
use in actual computations. However, for particular values of the
indice $p$, it can be reduced and written in a more compact form.
\begin{itemize}
\item For $p=1$, all the products in \eqref{formula} collapse and one
  is left with just an alternating sum:
  \begin{eqnarray}
    b_1^{(m,n)} & = & \sum_{\ell=0}^{m-1} Y_{n+\ell}^{(-1)^{m+1+\ell}}
    \\ & = & \sum_{\ell=0}^{m-1} 1 +(-1)^{m+1+\ell} \, k\nu \left(
    \frac{P_{n+\ell}}{n+\ell} + \frac{P_{n+\ell+1}}{n+\ell+1} \right)
    \\ & = & m + k\nu \left( (-1)^{m+1} \frac{P_n}{n} +
    \frac{P_{n+m}}{n+m} \right)\ ;
  \end{eqnarray}
\item For $p=m$, all the sums in \eqref{formula} collapse and one is
  left with:
  \begin{eqnarray}
    b_m^{(m,n)} & = & \prod_{\ell=0}^{m-1}
    Y_{n-\ell}^{(-1)^{m+1+\ell}} \\ & = & \prod_{\ell=0}^{m-1} \left(1
    - (-1)^{m+\ell} \frac{k\nu \left(n-\ell+ P_{n-\ell}
      \right)}{(n-\ell)(n-\ell+1)} \right) \\ & = &
    \prod_{\ell=0}^{m-1} \left(\frac{\left( n-\ell+ P_{n-\ell})(n-\ell
      + 1 - P_{n-\ell} -(-1)^{m-\ell} k\nu
      \right)}{(n-\ell)(n-\ell+1)} \right) \ .
  \end{eqnarray}
  One can then show, by considering separate cases for the parity of
  $m$ and $n$, the following identities:
  \begin{equation}
    \prod_{\ell=0}^{m-1} (n-\ell+1-P_{n-\ell}-(-1)^{m-\ell}k\nu) = 2^m
    \left[\tfrac{n-k\nu+(-1)^m P_{n+1}}2\right]_{\tfrac{m-P_m}2}
    \left[\tfrac{n+k\nu-(-1)^m P_{n+1}}2\right]_{\tfrac{m+P_m}2}
  \end{equation}
and   
  \begin{equation}
    \prod_{\ell=0}^{m-1}
    \frac{(n-\ell+P_{n-\ell})}{(n-\ell)(n-\ell+1)} =
    \frac{2^{-m}}{\left[\tfrac{n+P_{n+1}}2\right]_{\tfrac{m+P_m}2}
      \left[\tfrac{n-P_{n+1}}2\right]_{\tfrac{m-P_m}2}} \;,
  \end{equation}
which give, altogether:
  \begin{equation}
    b_{m}^{(m,n)} = \frac{\left[ \tfrac{n-k\nu+(-1)^m P_{n+1}}2
        \right]_{\tfrac{m-P_m}2} \left[ \tfrac{n+k\nu-(-1)^m P_{n+1}}2
        \right]_{\tfrac{m+P_m}2}}{\left[ \tfrac{n+P_{n+1}}2
        \right]_{\tfrac{m+P_m}2} \left[ \tfrac{n-P_{n+1}}2
        \right]_{\tfrac{m-P_m}2}}
    \label{last_b} \ ,
  \end{equation}
  where we used the \it descending Pochhammer symbol \rm defined as:
  $[a]_n = \frac{\Gamma(a+1)}{\Gamma(a-n+1)}, \ a \in \R, \, n \in
  \N$.
\end{itemize}

\paragraph{From the right.}
With again $n \geqslant m$, we can also look at what happens when the
monomial of highest degree is on the left --- and hence product have
to be performed toward the left\,\footnote{This difference between the
  structure constants produced when the highest degree monomial is
  placed on the left or on the right stems from the fact that, as
  explained in the Introduction, we choose to place the kleinian on
  the left of the resulting monomials.}. As we did previously, we
start by writing the expansion of the star product of two arbitrary
monomials as:
\begin{equation}
  (q_\a)^n \star (q_\b)^m = \sum_{p=0}^{m} \frac{(i\epsilon_{\a \b})^p
    \, n!}{(n-p)!} \, \bar b_p^{(n,m)} (q_\a)^{n-p} \, (q_\b)^{m-p} \ .
\end{equation}
Once again, a recurrence relation can be derived for these $\bar
b_p^{(n,m)}(\nu)$ coefficients:
\begin{eqnarray}
  (q_\a)^n \star (q_\b)^{m+1} & = & \Bigg(\sum_{p=0}^{m}
  \frac{(i\epsilon_{\a \b})^p \, n!}{(n-p)!} \, \bar b_p^{(n,m)}
  (q_\a)^{n-p} \, (q_\b)^{m-p} \Bigg) \star q_\b\nonumber \\ & = & \sum_{p=0}^{m}
  \frac{(i\epsilon_{\a \b})^p \, n!}{(n-p)!} \, \bar b_p^{(n,m)}
  \left[ (q_\a)^{n-p} (q_\b)^{m-p+1} + i\epsilon_{\a \b} (n-p) \bar
    Y_{n+m-2p} (q_\a)^{n-1-p} (q_\b)^{m-p} \right] \nonumber \\ & = &
  \sum_{p=0}^{m} \frac{(i\epsilon_{\a \b})^p \, n!}{(n-p)!} \, \bar
  b_p^{(n,m)} (q_\a)^{n-p} (q_\b)^{m-p+1} \nonumber \\ &
  & \hspace{75pt} + \sum_{p=1}^{m+1} \frac{(i\epsilon_{\a \b})^p \,
    n!}{(n-p)!} \, \bar b_{p-1}^{(n,m)} \bar Y_{n+m-2(p-1)}
  (q_\a)^{n-p} (q_\b)^{m-p+1} \nonumber \\ & = & \sum_{p=0}^{m+1}
  \frac{(i\epsilon_{\a \b})^p \, n!}{(n-p)!} \, \bar b_p^{(n,m+1)}
  (q_\a)^{n-p} \, (q_\b)^{m-p+1}\ ,
\end{eqnarray}
%
  \begin{equation}
    \Rightarrow \quad \bar b_p^{(n,m+1)} \;=\; \bar b_p^{(n,m)} + \bar
    Y_{n+m-2(p-1)} \, \bar b_{p-1}^{(n,m)}, \ \quad \forall \ p \leqslant m
    \leqslant n \in \N\, .
    \label{recurrence2}
  \end{equation}
%
Exactly as in the previous case, an exact expression for these
coefficients is given by sums of products of $\bar Y$'s:
%
  \begin{equation}
    \bar b_p^{(n,m)} \;=\; \prod_{\ell=0}^{p-1} \;\underset{i_0 =
      0}{\sum_{i_\ell = i_{\ell-1}}^{m-p}} \bar Y_{n+i_\ell - \ell} \ .
      \label{formulabis}
  \end{equation}
%
It can be checked that this expression solves the recurrence relation
given above, the proof being essentially the same as for the
coefficients $b_p^{(m,n)}$.

\paragraph{Reduced formula.}
Due to its structure, similar to \eqref{formula}, \eqref{formulabis} can
be simplified in two limit cases:
\begin{itemize}
\item For $p=1$, one is only left with a sum that
  can be performed explicitly:
  \begin{equation}
    \bar b_1^{(n,m)} = m + k\nu \left( \frac{P_n}{n} -
    \frac{P_{n+m}}{n+m} \right) \, ;
  \end{equation}
\item For $p=m$, only the product remains, which can also be computed,
  considering separate cases according to the respective parity of $m$
  and $n$:
  \begin{equation}
    \bar b_m^{(n,m)} = \frac{\left[ \frac{n+k\nu-P_{n+1}}2
        \right]_{\frac{m-(-1)^n P_m}2} \left[ \frac{n-k\nu+P_{n+1}}2
        \right]_{\frac{m+(-1)^n P_m}2}}{\left[ \frac{n-P_{n+1}}2
        \right]_{\frac{m-P_m}2} \left[ \frac{n+P_{n+1}}2
        \right]_{\frac{m+P_m}2}} \ .
  \end{equation}
\end{itemize}
%

\section{$\W$ algebras interlude} \label{sec:3}
Before relating the structure constants $b_p^{(2m,2n)}$ (of $\hsl$) to
those postulated in \cite{Pope:1989sr}, we succinctly recall
throughout the following section what are $\W$ algebras and their
relation to 3D higher spin algebras. For a more complete introduction
to $\W$ algebras, see e.g. \cite{Bouwknegt:1992wg}.\\

$\W_{N,c}$ algebras naturally appear in the context of 2D conformal
field theories involving higher spin currents
\cite{Zamolodchikov:1985wn}, i.e. with spins $s \in \llbracket 2, N
\rrbracket\,$, and central charge $c$.  They can be thought of as
higher spin extensions of the Virasoro algebra (describing a spin 2
quasi-primary current, namely the stress-energy tensor) in this
context. Because of the non linear terms appearing in the Operator
Product Expansion (OPE) of such higher spin currents, the structure of
the corresponding $\W_{N,c}$ algebras becomes quite intricate and, in
particular, they are not Lie algebras. \\

More recently, $\W$ algebras appeared as algebras of asymptotic
symmetries of 3D higher spin theories
\cite{Campoleoni:2010zq,Henneaux:2010xg,Henneaux:2012ny,Campoleoni:2011hg}.
When the higher spin theory involves an infinite tower of gauge fields
with all spin $s \in \N$, as in Prokushkin--Vasiliev's theory or the
Chern--Simons theory based on $\hsl \oplus \hsl\,$, the asymptotic
symmetry algebra is an infinite dimensional extension of the
$\W_{N,c}$ algebras, referred to as $\W_{\infty,c}\,$, which
corresponds to the algebra made out of all higher spin currents
together with the stress-energy tensor.  A first attempt to obtain
such an extension was carried out in
\cite{Pope:1989ew}\,\footnote{Actually the first appearance of such an
  higher spin extension was given in \cite{Bakas:1989xu}, which was
  later realised to be a particular contraction of the Pope, Shen and
  Romans $\Winf^{\PRS}$, denoted latter on as $w_\infty$.}  where the
authors obtained a Lie (hence linear) algebra, that we shall refer to
as $\Winf^{\PRS}$ hereafter, further explored in \cite{Pope:1990kc,
  Pope:1991ig}. In a later paper \cite{Pope:1989sr}, the authors
realised that there was a one-parameter family of such algebras, which
their first construction was a part of. They showed that for each
value of $\mu$ (a real number parametrising their family of
extension), these algebras admit a subalgebra, called \it wedge
subalgebra. \rm They found out that this subalgebra is isomorphic to
$\hsl\,$, where $\mu = \tfrac{\lambda^2-1}4$ is the value of the
quadratic Casimir of $\slr$, as both can be seen as the quotient
\eqref{hsl_def}. A puzzling feature of their construction for these
\it linear \rm infinite dimensional extensions of $\W_N$ algebras is
the fact that the introduction of an infinite number of generators
carrying negative spin is needed in order to satisfy the Jacobi
identity, except for the special value $\mu = 0\,$.  For this value,
the resulting, \emph{linear} infinite-dimensional algebra is the Lie
algebra denoted $\W^{\rm PRS}_{\infty}\,$.  For $\mu \neq 0\,$, it is
still not clear to us whether the resulting algebras containing
negative-spin generators can be related to \emph{nonlinear}
$\W_{\infty,c}\,$. \\

A more modern point of view in obtaining such extensions is given by,
from a mathematical point of view, the \it Drinfeld--Sokolov
reduction, \rm which associates to a semisimple Lie algebra a
centrally extended $\W$ algebra. This operation corresponds, from a
physical point of view \cite{Brown:1986nw}, to the passage of the
gauge algebra of some theory defined around anti-de Sitter (AdS)
background to its asymptotic symmetry algebra; see for instance
\cite{Gaberdiel:2011wb,Campoleoni:2011hg} for enlightening reviews of
the interplay between the two approaches.  The $\Winf$ algebras
obtained via the asymptotic symmetry algebra procedure do not suffer
from the odd feature of having negative spin generators, and, for
generic values of the parameter $\lambda$ of the higher-spin gauge
algebra $\hsl\,$, are nonlinear --- except for $\lambda=\pm 1$,
i.e. $\mu=0\,$ \cite{FigueroaO'Farrill:1992cv}.\\

In spite of the difficulties brought in by the appearance of nonlinear
terms, the structure constants of $\W_{\infty,c}$ have been derived
\cite{Campoleoni:2011hg} in terms of those of $\hsl\,$, which is
expected to coincide with its wedge in the limit $c\rightarrow
\infty\,$.  The structure constants of $\hsl$ were postulated in
\cite{Pope:1989sr}, in the Fourier basis given by the generators
$V^s_m$ carrying spin $s+2\,$, with $\rvert m \rvert \leqslant s+1\,$.
These generators verify:
\begin{equation}
  [V_m^i, V_n^j] = \sum_{\ell=0}^\infty g_{2\ell}^{i,j}(m,n)
  V_{m+n}^{i+j-2\ell}\;,
  \label{commutation_W}
\end{equation}
where $g_{2\ell}^{i,j}(m,n)$ are the structure constants given
hereafter. It was shown \cite{Pope:1989sr} that this commutation
relations could be realised as the antisymmetric part of an
associative algebra spanned by the same generators $V^s_m$ and endowed
with an associative product, the so-called \it ``lone-star
product''\rm:
\begin{equation}
  V_m^i \star V_n^j = \frac{1}2 \sum_{a=0}^\infty q^{a-1}
  g_{a-1}^{i,j}(m,n;\lambda) V_{m+n}^{i+j-a+1}\;,
  \label{lone_star}
\end{equation}
with
\begin{equation}
  g_a^{i,j}(m,n;\lambda) = \frac{1}{2(a+1)!} N_a^{i,j}(m,n)
  \phi_a^{i,j}(\lambda)\;,
  \label{w_struct1}
\end{equation}
\begin{equation}
  N_a^{i,j}(m,n) = \sum_{r=0}^{a+1} (-1)^r \binom{a+1}{r}
  [i+1+m]_{a+1-r} [i+1-m]_{r} [j+1+n]_{r} [j+1-n]_{a+1-r} \;,
  \label{w_struct2}
\end{equation}
and 
\begin{equation}
  \phi_a^{i,j}(\lambda) = {_4}F_3 \left[
    \begin{aligned}
      \tfrac{1-2\lambda}2, \ \, \tfrac{1+2\lambda}2, \ \,
      -\tfrac{a+1}2, \ \, -\tfrac{a}2 \\ -\tfrac{2i+1}2,
      -\tfrac{2j+1}2, i+j-a+\tfrac{5}2
    \end{aligned}
    ; 1 \right]\;.
  \label{w_struct3}
\end{equation}
The parameter $q$ can be used to rescale\,\footnote{It is in the limit
  $q\rightarrow 0$ that one recovers the algebra studied in
  \cite{Bakas:1989xu}.} the generators $V_m^s\,$.  The $\slr$ algebra
is generated by the $V^s_m$ with $s=0\,$, and its generators denoted
by $\{J_-, J_0,J_+\}$, obey 
\begin{equation}
  [J_+ , J_-] = 2 J_0\ , \quad [J_\pm , J_0] = \pm J_\pm \ ,
\end{equation}
with $J_0^\dagger = J_0$ and $J_\pm^\dagger = J_\mp\,$, in
accordance with $(V_m^s)^\dagger = V_{-m}^s\,$.

\section{Deriving the structure constants of $\shsl$} \label{sec:4}

In this section, we build the various powers of the $J_0$ generator of
$\slr$ out of deformed oscillators and prove that the structure
constants postulated in \cite{Pope:1989sr} are indeed those of $\hsl$,
or equivalently the wedge subalgebra of $\Winf\,$. We extend the result 
to the $\mathbb{Z}_2$-graded case of $\shsl\,$.
\subsection{Dictionary with the wedge subalgebra}
\noindent We define $w$ as
\begin{eqnarray}
  \label{number} 
  w = a^+ a^- = \tfrac{1}{2}\, \{ a^- , a^+\}_\star\ ,\qquad
  [w,a^\pm]_\star = \pm a^\pm\ ,
\end{eqnarray}
where $a^+$ and $a^-$ are the deformed creation and annihilation
operators, defined as
\begin{equation}
  a^\pm = u^{\pm \a}q_\a\ ,\quad u^{+\a} u^-_\a = \epsilon^{\a \b}
  u_\a^- u_\b^+ = -\frac{i}2\ ,\quad (u^{\pm}_\a)^\dagger =
  u^\mp_\a\ .
\end{equation}
These operators obey
\begin{equation}
  [ a^-, a^+]_\star = 1+\nu k\ ,\qquad \{k,a^\pm\}_\star=0 \ ,\qquad (
  a^\pm)^\dagger = a^\mp\ ,
  \label{dha}
\end{equation}
and can be used to realise $\slr$, by defining $J_0 = \tfrac{1}2 w$
and $J_\pm = \tfrac{1}2 (a^\mp)^2\,$.
The parameter $\nu$ of the deformed oscillators can be
related to $\lambda$, used to express the $\slr$ quadratic Casimir by
comparing \eqref{hsl_def} to the expression obtained when $\slr$ is
realised by these oscillators:
\begin{equation}
  \C[\slr]_{\text{osc.}} = \frac{1}{16} (k\nu - 3)(k\nu + 1) 
  = \frac{1}{16} (\nu^2 - 2 k\nu - 3) \, .
  \label{casimir_osc}
\end{equation}
At this point, further comments on the relation between $Aq(2;\nu)$
and $\hsl$ are in order. So far, we worked entirely in $Aq(2;\nu)\,$,
i.e. with arbitrary powers of the deformed oscillators $q$ and the
kleinian $k\,$.  As mentioned earlier, the associative algebra admits
a subalgebra $Aq(2;\nu)_e$ consisting of even-degree monomials in $q$,
together with powers $0$ or $1$ of $k\,$. This subalgebra can be
further decomposed into two consistent subalgebras, by projecting it
using $\Pi_\pm := \tfrac{1 \pm k}2\,$: $Aq(2;\nu)_e = \Pi_+
Aq(2;\nu)_e \oplus \Pi_- Aq(2;\nu)_e\,$
\cite{Vasiliev:1989re}. Therefore, working only in one of these two
projected subalgebras, the kleinian $k$ can be set to $\pm 1$.  As can
be seen from \eqref{def_gen_slr}, $k$ does not enter the construction
of the $\slr$ UEA from deformed oscillators, therefore to realise
$\hsl$ using them, we need to use only one of the two projections of
$Aq(2;\nu)_e$. The above relation \eqref{casimir_osc} then becomes $\C
= \tfrac{1}{16} (\nu^2 \mp 2 \nu - 3)\,$, which leads, upon comparing
it with \eqref{hsl_def}, to $\lambda = \tfrac{1\mp\nu}2\,$.
Hereafter, when we treat the bosonic algebra $\hsl\,$ we will work 
in the $\Pi_+$ projection of $Aq(2;\nu)_e$
i.e. we set $k\nu = \nu$, and $\lambda = \tfrac{1-\nu}2\,$. 
Otherwise, in the generic case of $\shsl\,$, we keep the Klein operator $k$
explicitly in the various expressions. \\

As a corollary of the formula \eqref{formula}, we can compute the star
product $w^m\star w^n\,$, where to fix the ideas we consider
$m\leqslant n\,$.  We start with
\begin{equation}
  (q_\a)^{2m} \star (q_\b)^{2n} = \sum_{p=0}^{2m}
  \frac{(i\epsilon_{\alpha \beta})^p \, (2n)!}{(2n-p)!} \,
  b_p^{(2m,2n)} (q_\a)^{2m-p}(q_\b)^{2n-p}
  \label{prod_even}
\end{equation}
and contract both sides with
\begin{align}
  (u^{+\alpha})^m (u^{-\alpha})^m (u^{+\beta})^n(u^{-\beta})^n\;.
\end{align}
This way, the left-hand side produces $w^m\star w^n\,$.  As for the
right-hand side, the structures that have a chance to give a nonzero
result when contracted with
$(i\epsilon_{\alpha\beta})^p\,(q_{\alpha})^{2m-p}\,
(q_\beta)^{2n-p}\,$ are
\begin{equation}
  \sum_{r=0}^{p} C^{(p,r)}_{(m,n)}\,\Big( \left[ (u^{+\alpha})^{r}
    (u^{-\beta})^{r}\right]
  \left[(u^{+\beta})^{p-r}(u^{-\alpha})^{p-r}\right]
  (u^{+\alpha})^{m-r} (u^{-\alpha})^{m-p+r} (u^{-\beta})^{n-r}
  (u^{+\beta})^{n-p+r}\Big)\;.
  \label{contraction}
\end{equation}
When $r=0$, one has the normalisation coefficient
$C^{(p,0)}_{(m,n)}=\frac{(2m-p)!(2n-p)! (p!)^2}{(2m)!(2n)!}
\binom{m}{p} \binom{n}{p}\,$. In the general case, one gets
$C^{(p,r)}_{(m,n)} = \frac{(2n-p)!}{(2n)!} \frac{(2m-p)!}{(2m)!} r!
(p-r)!  p! \binom{m}{r} \binom{m}{p-r} \binom{n}{r} \binom{n}{p-r}
\,$.
Using $[a]_n = \frac{a!}{(a-n)!} = \binom{a}{n} n!, \forall a \in \N$
such that $a \geqslant n$, one can rewrite $C^{(p,r)}_{(m,n)}$ as
follows:
\begin{eqnarray}
  C^{(p,r)}_{(m,n)} & = & \frac{(2n-p)!}{(2n)!} \frac{(2m-p)!}{(2m)!}
  r!  (p-r)!  p! \binom{m}{r} \binom{m}{p-r} \binom{n}{r}
  \binom{n}{p-r} \nonumber \\ & = & \frac{(2n-p)!}{(2n)!} \frac{(2m-p)!}{(2m)!}
  \binom{p}{r} [m]_r [m]_{p-r} [n]_r [n]_{p-r} \nonumber \\ & = &
  \frac{(2n-p)!}{(2n)!}  \frac{(2m-p)!}{(2m)!} \, c^{(p,r)}_{(m,n)} \ .
\end{eqnarray}
These coefficients obey $c^{(p,p-r)}_{(m,n)} = c^{(p,r)}_{(m,n)}\,$.
As a consequence, they obey $\sum_{r=0}^p (-1)^{r} c^{(p,r)}_{(m,n)} =
0$ for $p$ odd, and as a result, one obtains
%
  \begin{equation}
    w^m\star w^n = \sum_{p=0}^{2m}\left(\frac{1}{2}\right)^p
    \frac{(2m-p)!}{(2m)!}\; b_p^{(2m,2n)}\, \left[ \sum_{r=0}^p
      (-1)^{r} c^{(p,r)}_{(m,n)} \right]\, w^{m+n-p}\; \;,
    \label{w_product}
  \end{equation}
%
where only the even values of $p$ contribute. In the special case
$m=1$, we can use \eqref{last_b} to evaluate the above expression:
\begin{equation}
  w \star w^n = w^{n+1} - \frac{n^2}4
  \frac{(2n-1+\nu)(2n+1-\nu)}{(2n-1)(2n+1)} \ w^{n-1}\ ,
\label{wn}
\end{equation} 
which reproduces eq.\,(3.8) of \cite{Boulanger:2015uha}, where this
product was considered in the context of fractional spin gravity.\\

Using $V_0^0 = J_0 = \frac{1}2 w$, we are lead to the identification
$V_0^{n-1} = \left( \frac{1}2 \right)^n w^n$. To understand the origin
of this dictionary, let us clarify the difference between $w^n$ and
$\underbrace{w\star \dots \star w}_{n \text{ times}} \equiv w^{\star
  n}\,$.  The latter expression can be expanded in a sum of
Weyl-ordered monomials in the oscillators $q_\a\,$, starting with the
maximum degree $2n$ monomial corresponding to $w^n\,$, together with
lower-degree monomials. On the other hand, the generators $V_m^s$
being part of the enveloping algebra of $\slr$ can be expressed as
polynomials in $J_0$ and $J_\pm$.  In the envelopping algebra picture,
these generators are defined in terms of nested commutators of $\slr$
generators:\footnote{What we call here $V^{s-2}_m$ corresponds to
  $V^s_m$ in the conventions of \cite{Gaberdiel:2011wb}. There is
  therefore a shift of 2 units on the spin.}
\begin{equation}
  V_m^{s-1} := (-1)^{s-m}\, \frac{(s+m)!}{(2s)!} \,
  [\underbrace{J_-,[J_-,[\dots,[J_-}_{s-m \text{ times
          }},(J_+)^s]\dots]] = (-1)^{s-m}\, \frac{(s+m)!}{(2s)!} \,
    \left( \text{Ad}_{J_-} \right)^{s-m} \, \left( J_+ \right)^s
    \label{envelopping_def}
\end{equation}
which, upon using the commutation relations of $\slr$, can be reduced
to a polynomial in $J_0,\, J_\pm$. This is reminiscent of the fact
that $w^n$ is naturally expressed as a linear combination of ``star
power'' $w^{\star k}$, $0 \leqslant k \leqslant n$:
\begin{eqnarray}
  w^n & = & w \star w^{n-1} + \frac{(n-1)^2}{4}\,
  \frac{(2n-3+\nu)(2n-1-\nu)}{(2n-3)(2n-1)}\, w^{n-2} \nonumber \\ & = &
  w \star \left( w \star w^{n-2} + \frac{(n-2)^2}{4}\,
  \frac{(2n-5+\nu)(2n-3-\nu)}{(2n-5)(2n-3)}\, w^{n-3} \right) \\ & &+
  \frac{(n-1)^2}{4}\, \frac{(2n-3+\nu)(2n-1-\nu)}{(2n-3)(2n-1)} \,\left( w
  \star w^{n-3} + \frac{(n-3)^2}{4}\,
  \frac{(2n-7+\nu)(2n-5-\nu)}{(2n-7)(2n-5)} w^{n-4} \,\right) \nonumber
  \\ &=& \dots \nonumber
\end{eqnarray}
To sum up, 
\begin{itemize}
\item in the universal envelopping algebra picture, $(J_0)^n \equiv
  \underbrace{J_0 \otimes \dots \otimes J_0}_{n \text{ times}}$
  corresponds to taking $n$ star product of $\tfrac{1}2 w$,
  i.e. $(J_0)^n = \tfrac{1}{2^n} w^{\star n}\,$;
\item the symbol $w^n$ corresponds, in the envelopping algebra
  picture, to the $n$th power of the adjoint action of $J_-$ on the
  $n$th power of $J_+$, according to \eqref{envelopping_def}.
\end{itemize}
For instance, when $n=2$, we have:
\begin{equation}
  (\tfrac{1}{2} w)^2 = (\tfrac{1}{2} w) \star (\tfrac{1}{2} w) +
  \tfrac{1}{48} (1+\nu)(3-\nu) \equiv (\tfrac{1}{2} w) \star
  (\tfrac{1}{2} w) - \tfrac{1}3 \C[\slr]\ .
\end{equation}
Using \eqref{envelopping_def}, we can write:
\begin{equation}
  V_0^1 = \tfrac{1}{12} \, [J_-,[J_-,(J_+)^2]] = (J_0)^2 - \tfrac{1}3
  \C[\slr] \ .
\end{equation}
This reproduces the previous expression of $w^2$ upon making the
identifications $V_0^1 = \tfrac{1}4 w^2$ and $(J_0)^2 = \tfrac{1}4 w
\star w\,$, which motivates (and justifies) the previously proposed
relation $V_0^{n-1} = \left( \tfrac{1}2 \right)^n w^n\,$. \\

\noindent Using this dictionary, we can write:
\begin{eqnarray}
  w^m \star w^n & = & \sum_{p=0}^{2m} \left(\tfrac{1}2\right)^{p-1}
  g_{p-1}^{m-1,n-1}(0,0;\nu) w^{m+n-p} \nonumber \\ & = &
  \sum_{p=0}^{2m} \left( \frac{1}2 \right)^p \frac{(2m-p)!}{(2m)!}
  N_{p-1}^{m-1,n-1} b_p^{(2m,2n)}(\nu) w^{m+n-p}\ ,
\end{eqnarray}
which leads us to the following identification:
\begin{equation}
  b_p^{(2m,2n)}(\nu) = \binom{2m}{p}
  \phi_{p-1}^{m-1,n-1}(\nu)=:\binom{2m}{p} \Phi_{p}^{(m,n)}(\nu)\ ,
  \label{identification}
\end{equation}
where we used $N_{p-1}^{m-1,n-1} = \sum_{r=0}^p (-1)^r
c_{(p,r)}^{(m,n)}$. Let us emphasize that the sum \eqref{lone_star}
initially running over all integer values of $a$ has been truncated to
a finite sum from $0$ up to $2m$ (in the special case of interest to
us, i.e. where only generators $V_0^\ell$ are involved) because
$N_{p-1}^{m-1,n-1}$ vanishes for $p > 2m\,$. The reason behind this is
that for $p>2m$, either $[m]_r$ or $[m]_{p-r}$ vanishes for $r = 0,
\dots, p\,$. Recall that we supposed $m \leqslant n\,$, but had we
supposed the opposite, the same identification would have held and the
only change would be that the sum should run from $0$ up to $2n\,$.\\

In the special cases $p=1$ and $p=2m$, one can check that the previous
identification \eqref{identification} holds \,\footnote{Notice that it
  also holds for $p=0$, as the hypergeometric function then reduces to
  $1$.}, as it reproduces the formulas obtained previously from the
deformed star product. Indeed, $b_1^{(2m,2n)} = 2m$ as for $p=1$ the
last argument of the hypergeometric function (in the first row) is
zero, and therefore this function is equal to one. In the case $p=2m$,
the ${_4}F_3$ becomes a ${_3}F_2$ as the last argument in the first
row, $-(p-1)/2$ is equal to the first one in the second row,
$-m+1/2\,$. Then one can use Saalsch\"utz's theorem to evaluate it,
which yields:
\begin{equation}
  {_3}F_2 \left[
    \begin{aligned}
      \tfrac{\nu}2, \ \ \, 1-\tfrac{\nu}2, \ \ \, -m, \ \ \,
      \\ -n+\tfrac{1}2, n-m+\tfrac{3}2
    \end{aligned}
    ; 1 \right] = \frac{(\frac{1-2n-\nu}2)_m
    (\frac{\nu-2n-1}2)_m}{(\frac{1-2n}2)_m (-\frac{2n+1}2)_m} =
  \frac{[\frac{2n-1+\nu}2]_m [\frac{2n+1-\nu}2]_m}{[\frac{2n-1}2]_m
    [\frac{2n+1}2]_m}
\end{equation}
where $(a)_n$ is the raising Pochhammer symbol, i.e. $(a)_n =
\frac{\Gamma(a+n)}{\Gamma(a)}$, which obey $(-a)_n = (-1)^n [a]_n$
that we used in the second equality.\\

In order to prove that the structure constants of the lone
star-product \eqref{w_struct1} give the structure constants of
$\hsl\,$, we need to show that they obey the same recurrence relation
\eqref{recurrence} as the structure constant for the deformed
star-product.  In other words, we need to prove that the following
relation is true:
\begin{eqnarray}
  \label{hypergeometric_relation}
  \binom{2m+2}{p} \phi_{p-1}^{(m,n-1)} & = & \binom{2m}{p}
  \phi_{p-1}^{(m-1,n-1)} + 2 \binom{2m}{p-1} \phi_{p-2}^{(m-1,n-1)}
  \\ & & +
  \frac{(n+m-p+3/2+\nu/2)(n+m-p+5/2-\nu/2)}{(n+m-p+3/2)(n+m-p+5/2)}
  \binom{2m}{p-2} \phi_{p-3}^{(m-1,n-1)} \ , \nonumber
\end{eqnarray}
which is nothing but the recurrence relation \eqref{recurrence} that
is satisfied by the ``even-to-even'' coefficients $b_p^{(2m,2n)}\,$.

The relation \eqref{hypergeometric_relation} was proven in
\cite{Korybut:2014jza} (that we will not reproduce here), where a
different point of view from ours was adopted.  In our paper we
started by deriving a recurrence relation between the structure
constants of the star product between two monomials in the deformed
oscillators, i.e. the elements of $\hsl\,$.  We were also able to
compute explicitly the first and last of these coefficients, and match
them with the corresponding $\Winf^\PRS$ ones, upon identifying $w$
with $\tfrac{1}2 J_0\,$. Finally, the element of our proof is the
validity of \eqref{hypergeometric_relation}, showing that the
structure constants in the bosonic sector of $\hsl$ are those of the
wedge subalgebra of $\Winf^\PRS\,$. We end up with :
%
  \begin{equation}
    b_p^{(2m,2n)} = \binom{2m}{p} {_4}F_3 \left[
    \begin{aligned}
      \tfrac{\nu}2, \ \ \, 1-\tfrac{\nu}2, \ \ \, -\tfrac{p}2, \ \ \,
      -\tfrac{p-1}2 \ \, \\ -m+\tfrac{1}2, -n+\tfrac{1}2, n+m-p+\tfrac{3}2
    \end{aligned}
    ; 1 \right] \equiv \binom{2m}{p} \Phi_p^{(m,n)}(\nu)\ .
    \label{structure_even}
  \end{equation}
%
This hypergeometric function is in fact truncated because of the
Pochhammer symbol of the negative integer $-\tfrac{p}2$ or
$-\tfrac{p-1}2$. In can be rewritten as a finite sum, namely:
\begin{eqnarray}
  b_p^{(2m,2n)} & = & \binom{2m}{p} \sum^{[p/2]}_{k=0}
  \frac{(\tfrac{\nu}2)_k (1-\tfrac{\nu}2)_k (-\tfrac{p}2)_k
    (-\tfrac{p-1}2)_k}{k! (\tfrac{1-2m}2)_k (\tfrac{1-2n}2)_k
    (n+m-p+3/2)_k} \nonumber  \\ & = & \binom{2m}{p} \sum^{[p/2]}_{k=0}
  \frac{(\tfrac{\nu}2)_k (1-\tfrac{\nu}2)_k [\tfrac{p}2]_k
    [\tfrac{p-1}2]_k}{k! [\tfrac{2m-1}2]_k [\tfrac{2n-1}2]_k
    (n+m-p+3/2)_k}\ ,
\end{eqnarray}
where $[x]$ stands for the integer part of $x\,$.  From the
identification detailed in a previous paragraph between generators of
the wedge subalgebra of $\Winf^\PRS$ and monomials of even powers in
the deformed oscillators, we can infer that in this case, the
structure constants have the same expression whether the highest
monomial is on the left or the right side of the product, i.e. $\bar
b_p^{(2n,2m)} = b_p^{(2n,2m)}$. One way to check this is to look at
the recurrence relation linking the coefficients $\bar b_p^{(2n,2m)}$:
\begin{eqnarray}
  \bar b_p^{(2n,2m+2)} & = & \bar b_p^{(2n,2m+1)} + \bar
  Y_{2(n+m-p+1)+1} \bar b_{p-1}^{(2n,2m+1)} \nonumber \\ & = & \bar
  b_p^{(2n,2m)} + \left( \bar Y_{2(n+m-p+1)+1} + \bar Y_{2(n+m-p+1)}
  \right) \bar b_{p-1}^{(2n,2m)} + \bar Y_{2(n+m-p+1)+1} \bar
  Y_{2(n+m-p+2)} \bar b_{p-2}^{(2n,2m)} \nonumber \\ & = & \bar
  b_p^{(2n,2m)} + 2 \bar b_{p-1}^{(2n,2m)} +
  \frac{(n+m-p+3/2+\nu/2)(n+m-p+5/2-\nu/2)}{(n+m-p+3/2)(n+m-p+5/2)}
  \bar b_{p-2}^{(2n,2m)}\ ,
\end{eqnarray}
which is indeed the same recurrence relation as the one obeyed by
coefficients $b_p^{(2m,2n)}\,$. Added to the fact that the exact
expressions we obtained for $p=1$ and $p=m$ coincide with those of the
$b$-type coefficients, it justifies the equality $\bar b_p^{(2n,2m)} =
b_p^{(2m,2n)}$. Therefore, the coefficients entering the star product
of two monomials where the one of highest degree is on the left, and
an even power of the deformed oscillators, are:
\begin{eqnarray}
  \bar b_p^{(2n,2m)} & = & \binom{2m}{p} \Phi_p^{(m,n)}(\nu) =
  b_p^{(2m,2n)}\ .
\end{eqnarray}

\subsection{The $\mathbb{Z}_2$-graded case of $\shsl\,$}

We can now use the recurrence relation \eqref{recurrence} to derive
the coefficients $b_p^{(2m+1,2n)}$:
\begin{equation}
  b_p^{(2m+1,2n)} = \pi(b_p^{(2m,2n)}) +
  \frac{n+m-p+3/2+k\nu/2}{n+m-p+3/2} \, \pi(b_{p-1}^{(2m,2n)})\, .
\end{equation}
Finally, we need the coefficients in the case where $n$ is odd. Using
\eqref{right}, one can then show that the following recurrence
relation holds:
\begin{equation}
  b_p^{(m,n+1)} = \frac{n+1-p}{n+1}\; b_p^{(m,n)} +
  \frac{m+1-p}{n+1}\; \bar Y_{n+m-2(p-1)} b_{p-1}^{(m,n)}\, .
\end{equation}
This enables us to derive the remaining coefficients:
\begin{equation}
  b_p^{(2m,2n+1)} = \frac{2n-p+1}{2n+1}\; b_p ^{(2m,2n)} +
  \frac{2m-p+1}{2n+1}\; \frac{n+m-p+3/2-k\nu/2}{n+m-p+3/2}\;
  b_{p-1}^{(2m,2n)}\, ,
\end{equation}
\begin{eqnarray}
  b_p^{(2m+1,2n+1)} & = & \frac{2n-p+1}{2n+1} \;\pi(b_p^{(2m,2n)}) +
  2\frac{n+m-p+3/2+k\nu/2}{2n+1} \;\pi(b_{p-1}^{(2m,2n)}) \\ & & +
  \frac{2m-p+2}{2n+1}\; \frac{n+m-p+3/2+k\nu/2}{n+m-p+3/2}\;
  \frac{n+m-p+5/2+k\nu/2}{n+m-p+5/2}\; \pi(b_{p-2}^{(2m,2n)})\,
  .\nonumber
\end{eqnarray}
Turning now to the $\bar b$ coefficients, and using the egality $\bar
b_p^{(2n,2m)} = b_p^{(2m,2n)}$, we can deduce:
\begin{equation}
  \bar b_p^{(2n,2m+1)} = \bar b_p^{(2n,2m)} +
  \frac{n+m-p+3/2-k\nu/2}{n+m-p+3/2} \, \bar b_{p-1}^{(2n,2m)}\, .
\end{equation}
Finally, one can show the following other recurrence relation:
\begin{equation}
  \bar b_p^{(n+1,m)} = \frac{n-p+1}{n+1} \pi(\bar b_p^{(n,m)}) +
  \frac{m-p+1}{n+1} Y^+_{n+m-2(p-1)} \pi(\bar b_{p-1}^{(n,m)})\, ,
\end{equation}
yielding:
\begin{equation}
  \bar b_p^{(2n+1,2m)} = \frac{2n-p+1}{2n+1} \pi(\bar b_p^{(2n,2m)}) +
  \frac{2m-p+1}{2n+1} \frac{n+m-p+3/2+k\nu/2}{n+m-p+3/2}
  \pi(b_p^{(2n,2m)})\, ,
\end{equation}
\begin{eqnarray}
  \bar b_p^{(2m+1,2n+1)} & = & \frac{2n-p+1}{2n+1} \pi(\bar
  b_p^{(2m,2n)}) + 2\frac{n+m-p+3/2+k\nu/2}{n+m-p+3/2} \pi(\bar
  b_{p-1}^{(2m,2n)}) \\ & & + \frac{2m-p+2}{2n+1}
  \frac{n+m-p+3/2+k\nu/2}{n+m-p+3/2}
  \frac{n+m-p+5/2+k\nu/2}{n+m-p+5/2} \pi(\bar b_{p-2}^{(2m,2n)})\,
  . \nonumber
  \label{relation_bs}
\end{eqnarray}
Comparing these coefficients to the previous ones (the $b$-type ones),
one can notice the following relations:
\begin{equation}
  \begin{aligned}
    \bar b_p^{(2n,2m)} & = & b_p^{(2m,2n)}, & \ & \bar
    b_p^{(2n+1,2m+1)} & = & b_p^{(2m+1,2n+1)},
  \end{aligned}
\end{equation}
\begin{equation}
  \begin{aligned}
    \bar b_p^{(2n+1,2m)} & = & \pi(b_p^{(2m,2n+1)}), & \ & \bar
    b_p^{(2n,2m+1)} & = & \pi(b_p^{(2m+1,2n)})
  \end{aligned}
\end{equation}

\paragraph{Reconciling barred and unbarred coefficients.} 
A puzzling feature of the above derivation of the structure constants
for the free algebra generated by the deformed oscillator (modulo
their commutation relation) is the two types of coefficients that we
designated by $b_p^{(m,n)}$ and $\bar b_p^{(n,m)}$, distinguishing
between two situations: namely whether the lower degree monomial (of
deg. $m$) is on the left or the right side of the star product. This
distinction arises from the way we defined both kind of structure
constants, i.e. with all the lower order monomial in the oscillators
on the right side of those coefficients. Having in mind an ``operator"
form for the star product:
\begin{equation}
  f(q) \star g(q) = f(q) K(\overleftarrow \partial, \overrightarrow
  \partial, \overleftarrow \Delta, \overrightarrow \Delta, k\nu) g(q)
\end{equation}
with $K$ some function of the derivative and homogeneity operators
$\overrightarrow \Delta := q^\alpha
\frac{\overrightarrow\partial}{\partial q^\alpha}\,$, as well as of
$k\nu$, it seems more natural to look at the star product of two
monomial in the following form:
\begin{equation}
  q_{\a(m)} \star q_{\b(n)} = \sum_{p=0}^{\text{min(m,n)}}
  q_{\alpha(m-p)} \left( (i\epsilon_{\a\b})^p c_p^{(m,n)} \right)
  q_{\b(n-p)}\ ,
\end{equation}
where the $k$-dependent structure constants $c_p^{(m,n)}$ appear in
the middle, between two Weyl-ordered monomials.

The monomial $q_{\a(m-p)}$ and $q_{\b(n-p)}$ on the right hand side
are naturally interpreted as resulting from the action of $p$
derivatives acting both on the right and on the left. Thus, the above
equation can naturally be seen as the expansion of $K$ in power of the
derivative operator, and the coefficients $c_p^{(m,n)}$ as the
(possibly re-summed) action of the homogeneity operators (also
carrying various powers of $k\nu$) on the monomials. One could expect
the homogeneity operators acting on the left and on the right to be on
the same footing, in the same sense as for the derivatives: when
expanding $K$ it appears that each time a derivative acts on the left,
another one comes that acts on the right (we know it is true for the
standard Moyal star product, and because of its associative nature, we
can expect it will remain true for the deformed star product). Let us
try to see if at least some of these expectations are realised by
relating the coefficients $b_p^{(m,n)}$ and $\bar b_p^{(m,n)}$. To do
so, we can just rewrite the expansion of $q_{\a(m)} \star q_{\b(n)}$
as above (hereafter we will assume $m \leqslant n$):
\begin{equation}
  q_{\a(m)} \star q_{\b(n)} = \sum_{p=0}^m
  \frac{(i\epsilon_{\a\b})^p\, n!}{(n-p)!}  \, q_{\a(m-p)}
  \pi^{(m-p)}(b_p^{(m,n)}) q_{\b(n-p)}
\end{equation}
\begin{equation}
  q_{\a(n)} \star q_{\b(m)} = \sum_{p=0}^m
  \frac{(i\epsilon_{\a\b})^p\, n!}{(n-p)!} \, q_{\a(n-p)}
  \pi^{(n-p)}(\bar b_p^{(n,m)}) q_{\b(m-p)}\;.
\end{equation}
Before comparing $\bar c_p^{(m,n)} := \pi^{(m-p)}(b_p^{(m,n)})$ with
$\tilde c_p^{(n,m)} := \pi^{(n-p)}(\bar b_p^{(n,m)})$, let us prove
the useful identity:
\begin{equation}
  Y_{n+i}^{(-)^{i}} = \bar Y_{n+i}^{(-)^{n-1}},
\end{equation}
where we introduced the notation $\bar Y_n^{(-)^\ell} = 1 + (-1)^\ell
k\nu \left[ \frac{P_n}{n} - \frac{P_{n+1}}{n+1} \right]$.
\begin{proof}
  We can distinguish two different cases:
  \begin{itemize}
  \item $n+i$ is even, which implies $\bar Y_{n+i}^{(-)^{n-1}} = 1 +
    (-1)^n k\nu \frac{1}{n+i+1} = 1 + (-1)^i k\nu \frac{1}{n+i+1}$
    ($n+i$ being even, $n$ and $i$ have the same parity);
  \item $n+i$ odd, which implies $\bar Y_{n+i}^{(-)^{n-1}} = 1 +
    (-1)^{n-1} k\nu \frac{1}{n+i} = 1 + (-1)^i k\nu \frac{1}{n+i}$
    ($n+i$ being odd, $n$ and $i$ have opposed parity).
  \end{itemize}
  Therefore we have
  \begin{equation}
    \bar Y_{n+i}^{(-)^{n-1}} = Y_{n+i}^{(-)^{i}}\ .
  \end{equation}
\end{proof}
\noindent
Using this last identity, we can easily show that $\bar c_p^{(m,n)} =
\tilde c_p^{(n,m)} \equiv c_p^{(m,n)}\,$.
\begin{proof}
  We just have to apply the sign flips encoded in the action of $\pi$:
  \begin{eqnarray}
    c_p^{(m,n)} & = & \pi^{(m-p)}(b_p^{(m,n)}) = \prod_{\ell=0}^{p-1}
    \sum_{i_\ell = i_{\ell-1}}^{m-p} Y_{n+i_\ell-\ell}^{(-)^{i_\ell +
        \ell -(p-1)}} \nonumber \\ & = & \prod_{\ell=0}^{p-1} \sum_{i_\ell =
      i_{\ell-1}}^{m-p} \bar Y_{n+i_\ell-\ell}^{(-)^{n-p}} =
    \pi^{(n-p)}(\bar b_p^{(n,m)}) = \tilde c_p^{(n,m)}\ .
  \end{eqnarray}
\end{proof}
\noindent This enables us to ``reconcile" both types of structure
constants $b$ and $\bar b$ in only one:
\begin{equation}
  c_p^{(m,n)} = \prod_{\ell=0}^{p-1} \sum_{i_\ell =
    i_{\ell-1}}^{\text{min}(m,n)-p}
  Y_{\text{max}(m,n)+i_\ell-\ell}^{(-)^{i_\ell - \ell + p-1}}\ ,
\end{equation}
the remaining difference of status between the lower and the higher
degree of the two monomials being (a priori) due to the action of the
homogeneity operators that probably enters the above formula in a
re-summed form.

\subsection{Fourier mode basis of $\shsl$ and supertrace} 
%
It is common to present the $\mathcal N = 2$ super-$\Winf$ algebra, of
which $\shsl$ is the wedge subalgebra, in a way that makes the
$\mathcal N = 2$ super-multiplet structure explicit
(e.g. \cite{Romans:1991wi, Candu:2014yva}). The content of $\shsl$ in
terms of super-multiplet is as follows:
\begin{equation}
    \begin{Bmatrix}
    & 1 & \\
    3/2 & & 3/2\\
    & 2 &
    \end{Bmatrix}\, , \hspace{10pt}
        \begin{Bmatrix}
    & 2 & \\
    5/2 & & 5/2\\
    & 3 &
    \end{Bmatrix}\, , \hspace{10pt}
    \begin{Bmatrix}
    & 3 & \\
    7/2 & & 7/2\\
    & 4 &
    \end{Bmatrix}\, , \dots
\end{equation}
with each super-multiplet grouping generators of spins $s, s+1/2$ and
$s+1\,$, the spin-1 generator corresponding to $J\,$, see Section
\ref{sec:intro}.  In this section, we give the precise dictionary
between $\shsl$ generators in the Fourier mode basis (familiar in the
CFT context) and in terms of the deformed oscillators. Generically,
the spin $s$ generator is realised as a monomial of order $\ell =
2(s-1)$ in the deformed oscillators, and there are $2s$ modes for this
generator, labelled by $|m| \leqslant s-1$, such that the difference
between the number of the two different kind of oscillators ($q_1$ and
$q_2$) is $2m\,$; see e.g. \cite{Candu:2014yva}.

\paragraph{Bosonic sector.} We have established that
\begin{equation}
  V_0^s = \frac{1}{2^{s+1}} w^{s+1} \ .
\end{equation}
By rescaling the generators $V_m^s$ in order to reintroduce the
scaling parameter $q$ (that we set earlier to $1/4$ in order to make
contact with \cite{Boulanger:2015uha}), renamed hereafter $\gamma$ so
as to avoid any confusion with the deformed oscillators, one can
check, using \eqref{prod_even}, that the generators
\begin{equation}
  V_m^s = 2^{s-1} \gamma^s (a^+)^{s+1-m} (a^-)^{s+1+m}
  \label{PRS_def}
\end{equation}
indeed verify the $\Winf^{\PRS}$ commutation relations:
\begin{equation}
  \left[ V_m^s, V_n^{s'} \right]_\star = \sum_{p=0}^{\text{min}(s,s')}
  \gamma^{2p} \, g_{2p}^{s,s'}(m,n;\nu) \ V_{m+n}^{s+s'-2p}\ ,
\end{equation}
where $g_{2p}^{s,s'}(m,n;\nu)$ is given by \eqref{w_struct1},
\eqref{w_struct2} and \eqref{w_struct3}.\\

In order to present $\shsl\,$ in the way described above, we need to
redefine its generators in the Fourier mode basis as follows:
\begin{equation}
    T_m^{(s)\sigma} := \tfrac{1}2 (2\gamma)^{s-2} (a^+)^{s-1-m}
    (a^-)^{s-1+m} P_\sigma\; , \quad P_\sigma := \tfrac{1+\sigma
      k}{2}\; , \quad s \geqslant 2\;, \quad |m| \leqslant s-1\; .
\end{equation}
Splitting them into bosonic generators $V^{(s) \sigma}_m$ related to
the ones introduced in \eqref{PRS_def} by $s \rightarrow s-2$, and the
fermionic ones that we will rewrite:
\begin{equation}
  T_{m'}^{(s) \sigma} \equiv G_{m,\epsilon}^{(s) \sigma} := \tfrac{1}2
  (2\gamma)^{s-2} (a^+)^{s-1-(m-\epsilon/2)} (a^-)^{s-1+m-\epsilon/2}
  P_\sigma\; ,
\end{equation}
where we introduced the split $m'=m-\epsilon/2$ as for ferminonic
generators, both $s$ and $m'$ are half integer.
The maximal finite-dimensional subsuperalgebra contained in $\shsl\,$
is $\osp(2|2)\,$ and is spanned by the generators $T^{(s),\pm}_{m}$ in
the sector where $s\in\{ 1,3/2,2\}\,$.  Explicitly, in terms of the
generators $\{ Q^{(i)}_m\}\,$, $m=\pm {1}/{2}\,$, together with
$\{L_{-1},L_{0},L_{1}\}$ and $J\,$, where we choose $\gamma=1/2$ and
\begin{eqnarray}
  & Q^{(i)}_m :=
  \tfrac{1}{2}\,(u^{+\alpha})^{1/2-m}(u^{-\alpha})^{\tfrac{1}{2}+m}\;Q^{(i)}_\alpha\;,
  \quad m\in \{\tfrac{1}{2}\,,-\tfrac{1}{2} \} \;, & \\ & L_m := i
  (u^{+\alpha})^{1-m}(u^{-\alpha})^{1+m}\,T_{\alpha\alpha}\; , \quad
  m\in \{ - {1}\,, 0\,, {1}\} \; . &
\end{eqnarray}
By comparison with the presentation \eqref{osp22}, we have 
\begin{eqnarray}
 & \{ Q^{(i)}_m , Q^{(j)}_n\} = \sigma_3^{ij}\,L_{m+n}\,+
  \tfrac{m-n}{2}\, \tau^{ij}\,J \;, \quad\quad [L_m , Q^{(i)}_n ] =
  \tfrac{1}{2}\,(\frac{m}{2}+n)\, Q^{(i)}_{n-m}\;,\quad [J, Q^{(i)}_m]
  = \tau^{ij} Q^{(j)}_m \, .& \nonumber\\ & &
\end{eqnarray}

\paragraph{Fermionic sector.} 
Using the previous definitions, we arrive at 
\begin{equation}
  V_m^{(s),\sigma} \star G_{n,\epsilon}^{(s'),\sigma'} = \frac{1}2
  \sum_{p=0}^{\text{min}(2s-2,2s'-1)} \frac{1}{2\, p!} \, \gamma^{p-1}
  \tilde N^{s,s'}_{p, \epsilon}(m,n) \, \varphi_p^{(s,s')}(\nu)
  \ G_{n,\epsilon}^{(s+s'-[p+1]),\sigma'} \delta_{\sigma+\sigma',0}\ ,
\end{equation}
with
\begin{equation}
  \tilde N_{p,\epsilon}^{s,s'}(m,n) = \sum_{r=0}^p (-1)^r \binom{p}{r}
         [s-1-m]_r \, [s-1+m]_{p-r} \, [s'-1-n
           +\tfrac{\epsilon}2]_{p-r} \, [s'-1+n-\tfrac{\epsilon}2]_r
\end{equation}
and
\begin{equation}
  \varphi_p^{(s,s')}(\nu) = \frac{1}{2(s'-1)} \left( (2s-(p+1))
  \pi(\Phi_p^{(s-1,s'-3/2)}) + p \, \frac{s+s'-1-p+k\nu/2}{s+s'-1-p}
  \pi(\Phi_{p-1}^{(s-1,s'-3/2)}) \right)\ .
\end{equation}
Using \eqref{relation_bs}, one can show:
\begin{equation}
  \left[ V_m^{(s),\sigma}, G_{n,\epsilon}^{(s'),\sigma'} \right]_\star
  = \sum_{p=0}^{2(\text{min}(s,s')-1)} \frac{1}{4\, p!} \,
  \gamma^{p-1} \tilde N^{s,s'}_{p, \epsilon}(m,n) \,
  \left(\varphi_p^{(s,s')} -(-1)^p \pi(\varphi_p^{(s,s')}) \right) \,
  G_{n,\epsilon}^{(s+s'-[p+1]),\sigma'}\delta_{\sigma'+\sigma,0}\ .
\end{equation}
One the other hand,
\begin{equation}
  G_{m,\epsilon}^{(s),\sigma} \star G_{n,\epsilon'}^{(s'),\sigma'} =
  \sum_{p=0}^{2\,\text{min}(s,s')-1} \frac{1}{2\,p!} \, \gamma^{p} \,
  M_{p,(\epsilon,\epsilon')}^{s,s'}(m,n) \, \psi_p^{(s,s')}(\nu)
  \ V_{m+n-(\epsilon+\epsilon')/2}^{(s+s'-p), \sigma'}
  \delta_{\sigma+\sigma',0}\ ,
\end{equation}
with
\begin{equation}
  M_{p,(\epsilon,\epsilon')}^{s,s'}(m,n) = \sum_{r=0}^p (-1)^r
  \binom{p}{r} [s-1-m+\tfrac{\epsilon}2]_r \,
        [s-1+m-\tfrac{\epsilon}2]_{p-r} \,
        [s'-1-n+\tfrac{\epsilon'}2]_{p-r} \,
        [s'-1+n-\tfrac{\epsilon'}2]_r\,
\end{equation}
and
\begin{eqnarray}
  \psi_p^{(s,s')} & = & \frac{(2(s-1)-p)(2(s'-1)-p)}{4(s-1)(s'-1)} \,
  \pi(\Phi_p^{(s-3/2,s'-3/2)}) \nonumber \\ & & +
  \;2p\,\frac{(s+s'-p-3/2+k\nu/2)}{4(s-1)(s'-1)} \,
  \pi(\Phi_{p-1}^{(s-3/2,s'-3/2)}) \\ & & \hspace{10pt} +
  \frac{p(p-1)}{4(s-1)(s'-1)} \,
  \frac{(s+s'-3/2-p+k\nu/2)(s+s'-1/2-p+k\nu/2)}{(s+s'-3/2-p)(s+s'-1/2-p)}
  \, \pi(\Phi_{p-2}^{(s-3/2,s'-3/2)}) \;.\nonumber
\end{eqnarray}
Using $M_{p,(\epsilon,\epsilon')}^{s,s'}(m,n) = (-1)^p
M_{p,(\epsilon',\epsilon)}^{s',s}(n,m)$, one ends up with:
\begin{equation}
  \left\{ G_{m,\epsilon}^{(s),\sigma}, G_{n,\epsilon'}^{(s'), \sigma'}
  \right\}_\star = \sum_{p=0}^{2(\text{min}(s,s')-1)} \frac{1}{p!} \,
  \gamma^{p} \, M_{p,(\epsilon,\epsilon')}^{s,s'}(m,n) \,
  \psi_{p}^{(s,s')}(\nu)
  \ \left(V_{m+n-(\epsilon+\epsilon')/2}^{(s+s'-p),\sigma'} - (-1)^p
  V_{m+n-(\epsilon+\epsilon')/2}^{(s+s'-p),-\sigma'} \right)
  \delta_{\sigma+\sigma',0}
\end{equation}

\paragraph{Supertrace.}
Turning to the supertrace, it is easy to extract the maximal
contraction of the star-product of two monomials of same degree $n\,$,
thereby reproducing the formula given in
\cite{Vasiliev:1989re}. Denoting $F[f(q,k)] := f(0,k)\,$, one directly
obtains
\begin{eqnarray}
  F[(q_\alpha)^n \star (q_\beta)^m ] =
  \delta_{n,m}\,\epsilon_{\alpha_1\beta_1}\ldots
  \epsilon_{\alpha_n\beta_n} \; T_n(k,\nu)\;,
\end{eqnarray}
where
\begin{eqnarray}
  T_{2m+2}(k,\nu) &=& (-1)^{m+1}(2m+2)!\, (1 - \tfrac{k\nu}{2m+3}\,)(1
  + k\nu) \prod_{\ell = 1}^m (1 - \tfrac{\nu^2}{(2\ell + 1)^2}\,)\;,
  \nonumber \\ T_{2m+1}(k,\nu) &=& i (-1)^{m}(2m+1)!\, (1 + k\nu)
  \prod_{\ell = 1}^m (1 - \tfrac{\nu^2}{(2\ell + 1)^2}\,)\;.
\end{eqnarray}
This result agrees with $T_n(k,\nu) = i^n n! \, b_n^{(n,n)}$ $= i^n n!
\bar b_n^{(n,n)}$, as these coefficients are the same for $n$ and $m$
both even or odd. Indeed:
\begin{equation}
  \prod_{\ell=1}^m \left( 1 - \frac{\nu^2}{(2\ell+1)^2} \right) =
  \prod_{\ell=1}^m \frac{\left( \ell + \frac{1+\nu}2 \right) \left(
    \ell + \frac{1-\nu}2 \right)}{\left( \ell +\frac{1}2 \right)^2} =
  \frac{(\tfrac{3-\nu}2)_m \, (\tfrac{3+\nu}2)_m}{((\tfrac{3}2)_m)^2},
\end{equation}
\begin{equation}
  \begin{aligned}
    b_{2m+1}^{(2m+1, 2m+1)} = \frac{(\tfrac{1+\nu}2)_{m+1}
      (\tfrac{3-\nu}2)_{m}}{(\tfrac{1}2)_{m+1} (\tfrac{3}2)_{m}} =
    (1+\nu) \frac{(\tfrac{3+\nu}2)_{m}
      (\tfrac{3-\nu}2)_{m}}{((\tfrac{3}2)_{m})^2}, & \ & b_{2m}^{(2m,
      2m)} = \frac{(\tfrac{1+\nu}2)_{m}
      (\tfrac{3-\nu}2)_{m}}{(\tfrac{1}2)_{m} (\tfrac{3}2)_{m}}
  \end{aligned}
\end{equation}
\begin{equation}
  \Rightarrow T_n(k,\nu) = i^n n! \, b_n^{(n,n)}\ .
\end{equation}
The supertrace Str${}_{\nu}[f(q,k)] = f(0,-\nu)$ can readily be
obtained from this result.  What is obvious from this formula is that
fact that the supertrace degenerates for critical values of
$|\nu|\,$. \\

Another special case (refered to as ``hypercritical'' in
\cite{Boulanger:2015uha}) is $\rvert \nu \rvert = 1$: because of the
factor $(1+k\nu)$ (upon setting $k=\pm 1$), the supertrace degenerates
and becomes identically zero. This is a well known feature of $\hsl$
at $\lambda=\pm1$.  In this case, $\mu=0$ which means that $\hsl\,$,
as an associative algebra, is the universal enveloping algebra of
$\slr$ quotiented by $\langle \C \rangle$, i.e. where the identity
operator is removed from the UEA. As a result, the invariant bilinear
trace being defined as taking the identity component of the product of
two elements, is degenerate. To circumvent this problem, one usually
rescales the trace by $\tfrac{1}{\lambda^2-1}$, or equivalently in our
case, by $\tfrac{1}{1\pm\nu}\,$.

\section{Conclusion}
Using the realisation of $\shsl$ provided by the associative algebra
made out of all \textit{symmetrised} (even and odd) powers of the
deformed oscillators endowed with the star-commutator defined by
\eqref{WeylOrder}, we have given closed-form formulae for the
structure constants of $\shsl\,$.  In particular, in the bosonic case
we gave a formal proof that the structure constants postulated in
\cite{Pope:1989sr} are indeed those of $\hsl\,$, thereby completing
the work of \cite{Korybut:2014jza}. Our proof relies on the
associativity of the deformed star product, from which follows the
recurrence relation \eqref{recurrence} (resp. \eqref{recurrence2})
linking the structure constants $b_p^{(m,n)}$ (resp. $\bar
b_p^{(m,n)}$) involved in the product of two monomials of degree $m$
and $n$ (resp. $n$ and $m$) to those for two monomials of degree $m+1$
and $n$ (resp. $n$ and $m+1$).
We were able to give closed-form formulae, \eqref{formula} and
\eqref{formulabis}, for the structure constants verifying the
aforementioned recurrence relation.  These formulae are expressed in
terms of nested sums of products of some elementary building blocks,
the linear functions denoted $Y_n^\pm$ and $\bar Y_n\,$, see
\eqref{nest1} and \eqref{right}.  The latter functions, via their
dependence in $\nu\,$, encode the deformation appearing in the star
product of a single oscillator with an arbitrary Weyl-ordered monomial
in the oscillators when one replaces the non-deformed oscillators
(with $\nu=0$) by the deformed ones.
Finally, we were able to show, in the bosonic case, that:
\begin{itemize}
\item[(1)] our closed formula agrees with the structure constants
  $g_p^{m,n}(\nu)$ of \cite{Pope:1989sr} for the first two and the
  last terms ($b_0^{(m,n)},\, b_1^{(m,n)}$ and $b_m^{(m,n)}$)
  appearing in the expansion of the star product in terms of pointwise
  products of lowest degree monomials;
\item[(2)] the structure constants involved in the star product of
  even monomials $b_p^{(m,n)}$ and $\bar b_p^{(n,m)}$, and the
  structure constants $g_p^{m,n}(\nu)$, verify the same recurrence
  relation (using a result of \cite{Korybut:2014jza}).
\end{itemize}
What the recurrence relations \eqref{recurrence} and
\eqref{recurrence2} show is that knowing the ``boundary data''
\begin{equation}
  \left\{b_0^{(m,n)},\, b_m^{(m,n)}, \ \forall m \leqslant n \in \N
  \right\}
\end{equation}
is sufficient to reconstruct any of the structure constants. As those
of \cite{Pope:1989sr} verify the two conditions enumerated above, they
therefore are the unique solution of \eqref{recurrence} and
\eqref{recurrence2}, and as a consequence, the Lone Star product
constructed in \cite{Pope:1989sr} is the deformed star product
\eqref{WeylOrder}.\\

\section*{Acknowledgments}
We want to thank Fabien Buisseret for collaboration at the beginning
of the project.  It is a pleasure to thank Andrea Campoleoni, Shouvik
Datta and Tom\'a\v{s} Proch\'azka for discussions on $\W$ algebras and
asymptotic symmetries, as well as Slava Didenko, Zhenya Skvortsov,
Philippe Spindel, Per Sundell and Mauricio Valenzuela for discussions
on the deformed star product.  T.B. also thanks Kevin Morand for
various discussions on the construction of $\hsl$ from deformed
oscillators.  T.B. is supported by a joint grant ``50/50''
Universit\'e Fran\c{c}ois Rabelais Tours -- R\'egion Centre / UMONS.

\newpage

\providecommand{\href}[2]{#2}\begingroup\raggedright\endgroup

\end{document}